# MACHINE LEARNING FOR PROTEINS' FUNCTIONS

# למידה ממוחשבת לפיענוח תפקודי חלבונים

Dan Ofer

# DECLARATION

This dissertation is the result of my own work and includes nothing, which is the outcome of work done in collaboration except where specifically indicated in the text. It has not been previously submitted, in part or whole, to any university of institution for any degree, diploma, or other qualification.

Signed:______Dan Ofer  דן עופר_____________

Date:_18.11.2015_

# SUMMARY

> *My original research contributions include: Creating frameworks for using machine learning on whole protein sequences. Novel methods for extracting alignment free features solely from primary sequences. The first automatic Neuropeptide precursor discovery system. A generic machine learning methodology for high level protein function.*

Systematic identification of protein function is a key problem in current biology. Most traditional methods fail to identify functionally equivalent proteins if they lack similar sequences, structural data or extensive manual annotations. I focused on identifying diverse classes of proteins that share functional relatedness but little sequence or structural similarity, notably, Neuropeptide Precursors (NPPs).

Neuropeptides (NPs) are short secreted peptides produced in neuronal cells via cleavage from longer NPPs. They may act as neuromodulators and neurohormones, with many roles in cellular communication and regulation. NPs act via GPCRs, by activating signaling cascades governing broad functions such as metabolism, sensation and behavior in all Metazoans. Existing alignment based tools fail to identify them, due to their diversity and length.

I aim to identify functional protein classes solely using unannotated protein primary sequences from any organism. This task required the identification of biologically and statistically meaningful characteristics amenable to statistical learning, which we refer to as features. This thesis focuses on feature representations of whole protein sequences, novel sequence derived engineered features, their extraction, the building of frameworks for their usage by statistical learning ("machine learning", ML) models, and the application of ML models to biological tasks, focusing on high level protein functions.

I implemented the ideas of feature engineering to develop a platform (called NeuroPID) that extracts meaningful features for successful classification of both known and overlooked NPPs using machine learning. The platform allows mass discovery of new NPs and NPPs. It was further expanded as a freely available webserver.

I expanded our computational approach towards identifying other challenging protein groups with alignment-free statistical features. This is implemented as a novel bioinformatics toolkit called ProFET (Protein Feature Engineering Toolkit). ProFET extracts hundreds of biophysical and sequence derived attributes, allowing the application of machine learning methods to any set of proteins.

ProFET was applied on many protein benchmark datasets with state of the art performance. The success of ProFET applies to a wide range of high-level functions such as metagenomic analysis, subcellular localization, structural fold and unique functional properties (*e.g.* thermophiles, nucleic acid binding). The ML approach is part of a growing body of methodologies that complement classical sequence-alignment based approaches to protein family classification.

The platforms that I developed for this thesis allow the use of a holistic machine learning approach to the pressing problem of protein function prediction. These methods and frameworks represent a valuable and implemented resource for using ML and computational data science methods on proteins.

The work presented here is based on a number of my peer-reviewed research articles (Ofer D *et al.*, 2014; Karsenty *et al.*, 2014; Ofer and Linial, 2015), as well as unpublished work.

# תקציר:

*ממצאי המחקר שלי הינם: שיטות חישוביות לכימות ואפיון רצפי חלבונים באמצעות למידה חישובית ללא מאגרי מידע חיצוניים. המערכת האוטומטית הראשונה לזיהוי פרקורסורים של נוירופפטידים. מערכות לעיבוד ואפיון חלבונים דרך למידה חישובית. שיטות חדשות לאפיון תכונות כמותיות על מנת לסווג תפקודי חלבונים משונים או מופשטים, אשר שיטות קלאסיות לא עבדו עבורם.*

הבנת תפקודי החלבונים הרבים המתגלים כיום הינה בעיה מרכזית בביולוגיה. בעבר סיווג תפקודי חלבונים התבסס על בדיקה ידנית ודמיון רצפי לפי עימוד רצפים. במקרים מרובים השיטות הקלאסיות אינן עובדות. למשל, במקרים בהם חסר דמיון רצפי או מבני בין חלבונים שונים בעלי פונקציה ייחודית ומשותפת. בעיה זו מוקצנת עבור זיהוי פונקציות מופשטות – למשל זיהוי חלבונים השייכים למשפחה רחבה לעומת זיהוי הומולוג קרוב מתת-משפחה צרה. דוגמה שחקרנו הינם הפרקורסורים של נוירופפטידים.

נוירופפטידים הם חלבונים קצרים מופרשים, הנוצרים עקב חיתוך חלבוני פרקורסור ארוכים יותר. הם משמשים לתקשורת בין-תאית במערכת העצבים. הפפטידים המופרשים פועלים על ידי קשירת רצפטור והפעלת תגובות שניוניות בתאי מטרה. הם נמצאים בכל בעלי החיים והם מעורבים במגוון עצום של מנגנונים ביולוגים: התנהגות חברתית, רבייה, מטבוליזם, חישה ועוד. כלים חישוביים קיימים המתבססים על עימוד רצפים אינם מצליחים לזהות נוירופפטידים חדשים, בשל חוסר דמיון רצפי ואורכם הקצר.

בניתי מספר כלים שנועדו לפתור את הבעיה הזו, יחד עם חלבונים מרובים אחרים אשר להם פונקציות אופייניות שאינן נחשפות ע"י כלים קיימים.
פיתחתי את NeuroPID. זוהי פלטפורמה חישובית לגילוי נוירופפטידים ופרקורסורים חדשים. האפיון נעשה על פי תכונות כמותיות שאנו שולפים ישירות מרצפי החלבונים. אנו משתמשים בתכונות לאימון וכוונון מודלים פרדיקטיביים. כמו כן בפרויקט בנינו אתר שמלווה כלי זה, לצורך שימוש נוח לכלל הקהילה המדעית.
בנוסף, הרחבתי את הכלים והשיטות הסטטיסטיות בהם נעזרתי לאפיון נוירופפטידים לאפיון כלל המשפחות, תפקודים אפשריים וסוגי חלבונים. כל זה נאסף בתור ProFET (Protein Feature Engineering Toolkit). יצרתי בזאת שיטות חדשות לייצוג חלבונים שלמים כאוסף "פיצ'רים" ומערכת לייצר ולכמת את הפיצרים מרצפים חדשים ולאפיינם.

הכלים ושיטות שפיתחתי במהלך המחקרים הללו מאפשרים את השימוש של שיטות למידה חישובית (ML) לבעיות המרובות וקשות של אפיון חלבונים ותפקודם. שיטות ML באות לחזק את היכולת שלנו ללמוד ולזהות תכונות חמקניות, רחבות היקף

או מאתגרות, אשר השיטות הקלאסיות לא מסוגלות לתפוס. הכלים הללו נועדו לאפשר שימוש נרחב של למידה חישובית על ידי הקהילה המדעית, לצורך הבעיות המרובות של אפיון חלבונים ותפקודם.

חומר התיזה מבוסס על המחקר שלי, בפרט על מספר מאמרים שפורסמו (Ofer D *et al.*, 2014; Karsenty *et al.*, 2014; Ofer and Linial, 2015), יחד עם חומר שעדיין לא פורסם.

# ACKNOWLEDGEMENTS

First, I would like to thank Prof. Michal Linial. Without her I would never have entered the field, or encountered the myriad opportunities it led me to. She gave me the opportunity to try and learn, despite my beginning as a Tabula-Rasa. She was available for fascinating discussions around the clock. She always knew to see to the core of any matter, and taught me to always be aware of wider scientific meaning and purpose even while probing away in research.

I thank the mentors and professors I have had the opportunity to learn from over the years, and who gave me a chance.

I thank my friends for support, advice, feedback and comradery.

I thank the many authors whose works inspired me – Asimov, Tolkien, Pratchett, Dawkins, Moore, Gaiman, Herbert, Clarke and dozens more.

I am thankful to all the members & friends of the lab, past and present. For helping when I was stuck, for assisting when I was at a loss, for giving ideas and sharing time and advice. Thanks Nadav R., Nadav B., Kerem, Shelly, Ohad, Roni, Amos & Nati.

Last but not least, I thank my wonderful family for endless support, love and encouragement. I thank my parents, sisters, and grandparents. I would not be around without them.

# CONTENTS

# LIST OF TABLES

# LIST OF FIGURES

# List of Abbreviations and Acronyms

AA, Amino Acids

BLAST, basic local alignment search tool

PSI-BLAST, Position-Specific Iterated BLAST

CAFA, critical automatic functional annotation

CASP, critical Assessment of protein Structure Prediction

ClanTox, classifier of animal toxins

CNS, central nervous system

ICI, ionic channel inhibitor

MS, mass spectrometry

nAChR, nicotinic acetylcholine receptors

NP, neuropeptide

NPP, neuropeptide precursor polypeptide

PSSM, position specific scoring matrix

SP, signal peptide

TMD, Trans-membrane domain

TOLIP, toxin-like proteins

GPCR, G-protein-coupled receptors

PC, Prohormone Convertases

SWP, Swiss-Prot

SVM, support vector machine

SVC, support vector machine classifier

RF, Random Forests

GBR, Gradient Boosting Trees

RBF, Radial Basis Function (kernel)

CV, Cross validation

HMM, Hidden Markov Model

ANN, Artificial neural network

ROC, receiver operating characteristic

AUC, Area under ROC curve

ML, Machine learning

NN, Nearest-neighbours

F1, Balanced F1 score (mean of precision and recall)

TP, True Positives

TN, True Negatives

FP, False Positives

FN, False Negatives

# 1 INTRODUCTION רקע מדעי

Recent years have seen an exponential increase in biological data, particularly of gene and protein sequences (Figure 1). Modern Next Generation Sequencing (NGS) technologies have created a reality where the extraction of biomolecular information (i.e their genome, proteome, transcriptome, peptidome, etc') is an increasingly routine act for individual researchers, thanks to the plummeting costs of sequencing. This revolution is also leading to increasing diversity in what is being studied, with non-model organisms being sequenced, resulting in an explosion of novel proteins with unknown functions. Understanding protein function is a foundational issue in biology. While expert manual annotation might have sufficed in the past, when most research was limited to a handful of intensely studied, often similar, organisms (e.g. Humans & lab rats), this approach fails to scale to the modern "big data" explosion of newly discovered proteins. The flood of high-throughput and rapidly sequenced proteomic data cannot be handled by performing individual experimental studies to determine the function (s) of every single protein, as was traditionally the case. Computational prediction is currently the only feasible approach for "unlimited"-throughput identification of protein function (Loewenstein *et al.*, 2009) The only feasible way to assign functions to this deluge of sequences is with state-of-the-art, scalable computational methods for automated functional annotation and analysis.

An additional, complementary task in the context of labelling protein functions is finding proteins which share various novel functions of interest (as opposed to annotating a given sequence with a certain function), e.g. discovering new neuronal receptors, antimicrobial activity, chaperones or secreted peptides. Thus, the two tasks are finding the functions of a sequence and finding sequences which share a function.

Overcoming the growing gap between unknown and annotated sequences requires improving and expanding automated methodologies; Improving means better predictive power, identification of functions that are currently neglected (e.g. "high level" functions (Varshavsky *et al.*, 2007), or toxin-like short proteins (Tirosh *et al.*, 2013; Naamati *et al.*, 2009)), while expansion necessitates wider participation by diverse experts in the scientific community, lowering barriers of entry and incorporating the knowledge of non-computational biologists. The most used approach in protein classification relies on distance measures between sequences according to sequence-alignment methods (e.g., Smith-Waterman (Smith and Waterman, 1981), BLAST (Basic local alignment search tool Altschul et al., 1990)). The approach is one of "guilt by association": If sequence A has function X, and sequence B is sufficiently similar to sequence A, then sequence B is annotated with A's functions. Similarity may be based on parts of a sequence, e.g. specific domains within a protein or short motifs (e.g. conserved binding sites). With the growth in the amounts and diversity of protein sequences, more sophisticated methods have been introduced (e.g., PSSM, HMM-HMM) (Jaakkola *et al.*, 2000; Biegert and Söding, 2009; Söding, 2005) . These methods use multiple sequence alignments (MSA), improving sensitivity in detecting remote homologues (Edgar and Sjölander, 2004; Karplus *et al.*, 1998). 3D-structure can serve to define seeds for defining statistical models further improving the quality of protein domains and families (*e.g.,* Pfam) (Finn *et al.*, 2014), but this is severely limited due to the limitations of structural prediction (Berg *et al.*, 2002). InterPro (Mulder and Apweiler, 2007) has ~27,000 such models, covering 83% of all sequences in UniProtKB (2014_10). Function assignment is gained from mapping InterPro models to Gene Ontologies (*i.e*. InterPro2GO). Alternative model-free approaches have been proposed (Portugaly *et al.*, 2002). Such approaches require alternate representations of the sequences as sets of features, followed by fitting ML models. The assessment of large-scale automatic protein functional annotations (e.g. CASP, CAFA) (Radivojac *et al.*, 2013) and the contribution of alternative approaches has been discussed extensively (Juncker *et al.*, 2009; Yachdav *et al.*, 2014; Obradovic *et al.*, 2005).

Despite the specialized strengths of the sequence-alignment/model-based methods, in many cases they fail at reliably assign function (Rentzsch and Orengo, 2009). This is best demonstrated by their limitations in classifying 3D structural folds (Dubchak *et al.*, 1995; Ding and Dubchak, 2001; Greene *et al.*, 2007; Kister and Potapov, 2013), a task at which even the best sequence similarity based methods underperform. Many biological niches and functional properties are especially unsuited to this form of representation. For example, routine annotation tools fail to confidently assign function to bioactive peptides and short proteins (Naamati *et al.*, 2009).

A number of previous studies focused on features extracted from whole protein sequences, (Nanni *et al.*, 2010; Dobson *et al.*, 2009; Scheraga and Rackovsky, 2014; Dubchak *et al.*, 1995; van den Berg *et al.*, 2014; Nanni *et al.*, 2014; Varshavsky *et al.*, 2007) as a starting input for machine learning methodologies. Structural benchmarks (3D structural classes, folds or super-families) from SCOP and CATH (Greene *et al.*, 2007; Andreeva *et al.*, 2014) were used to assess many of these. Specialized predictors have been presented for a wide range of structural and sequence-property tasks including secondary structure, solvent accessibility, disordered regions, domains and more (C. . Z. Cai *et al.*, 2003; Ding and Dubchak, 2001; Cheng and Baldi, 2006; Deng *et al.*, 2015; Nagarajan and Yona, 2004; Barla *et al.*, 2008). ML approaches have proven suitable to classify many protein properties beyond 'just' structure, e.g. SVMProt was tested on 50 Pfam families (C. Z. Z. Cai *et al.*, 2003). Feature based ML classification outperformed alignment-based methods for many protein families (Varshavsky *et al.*, 2007). Alignment based distance metrics can even be incorporated as additional features (Liu *et al.*, 2014). However, the most likely advantage of the feature and pattern-based ML approach is in learning high-level functions (e.g. ProFET (Ofer and Linial, 2015)). Examples for such include protein-protein interactions (Bock and Gough, 2001; Cheng and Baldi, 2007), discriminating outer membrane proteins (Gromiha and Suwa, 2005), membrane topology (Nugent and Jones, 2009), and subcellular localization (Petersen *et al.*, 2011; Yachdav *et al.*, 2014; Goldberg *et al.*, 2014). However, most ML protein sequence analysis studies in the field focused on residue (rather the whole-sequence) level properties, such as secondary structure, disorder or PTMs such as proteolytic cleavage. The design of feature representations for residue level properties is far easier conceptually, as features are extracted from fixed length regions ("windows") rather than variable length whole proteins, e.g. encoding the amino acids in each window position as a feature. I would argue that this helps explains the popularity of ML methods in residue level predictive tasks, as opposed to whole protein level prediction.

This background serves as a basis for the approach I took in creating a ML platform for predicting high level protein functions (including NPPs).

We illustrate the significance of machine learning tools in identifying and annotating short bioactive proteins and peptides from many genomes. Over 5,700,000 full-length proteins from metazoans are currently archived, and some ~15% are short or secreted proteins (Figure 2). Most short sequences remain uncharacterized. Furthermore, due to incomplete genomes and transcriptomes, many sequences are actually fragments.

The systematic discovery of overlooked secreted peptides, neuropeptides and toxin-like proteins allows for new strategies in research and behavioural manipulation, shining a light on peptides as biologically and clinically important entities.

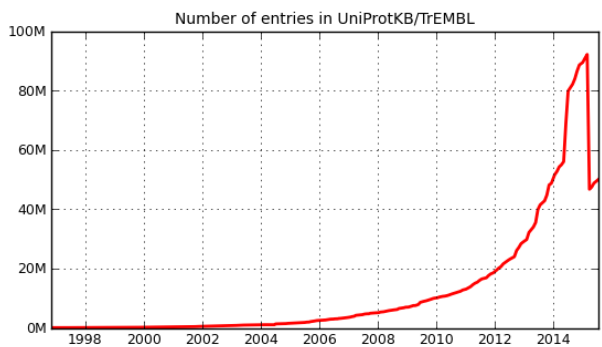

**Figure 1 Growth in sequenced proteins. Source: UniProtKB Release Statistics 2015_08.**

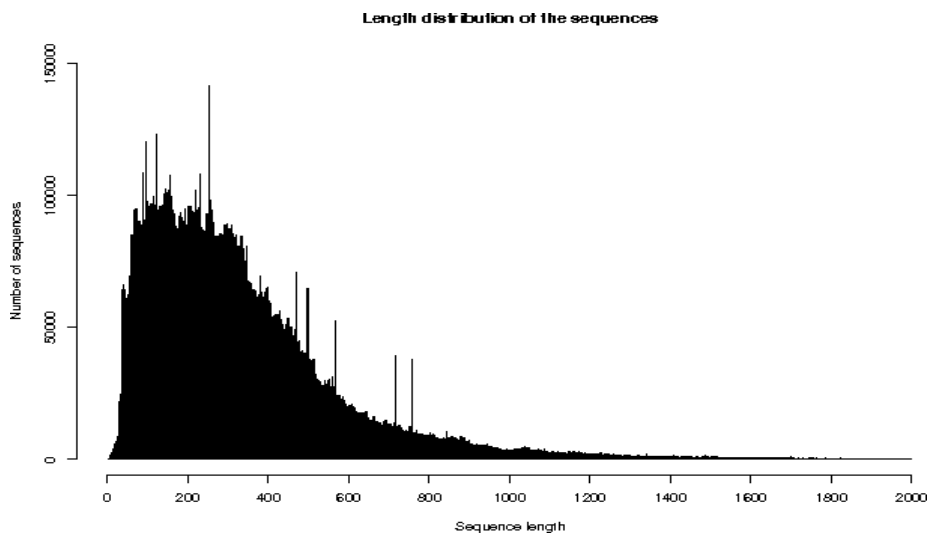

Figure 2   Distribution of sequence lengths in the UniProtKB database (excluding fragments). Lengths range from 2 to 35,213, Average: 356. Distribution covers 46 Million sequences. Source: UniProtKB Release Statistics 2015_08.

## 1.1 Machine Learning & Data Science

I describe the steps I applied during my research and will share the logic and concepts fundamental to assessing success and addressing the challenges inherent to applying ML tools to proteins. To implement any supervised ML task, a number of steps must be performed, regardless of the field. These include:

### 1.2.1 Data Collection

Data must be defined and gathered as a dataset. This involves the curation, acquisition and sometimes creation of the data, samples and measurable parameters. In the case of proteins, this means acquisition of the protein sequence, relevant annotations (e.g. class labels, subcellular localizations, structural fold, genomic data), storage in machine-readable formats and data cleaning ("munging"), e.g. removal of unreliable sequences, errors, and inaccurate labels.

### 1.2.2 Feature Extraction

Raw data must be transformed into a quantitative, computable representation. It must be amenable to representation as a fixed length vector of features, as the vast majority of ML methods require a fixed length input (of any arbitrary length). This is a completely different task for every domain: in the case of computer vision, it means extracting raw pixel values (from images or "crops" of identical sizes). In the case of stocks or Electroencephalograms, it involves extracting time-series and amplitudes. In the case of natural text, it may involve counting the words or characters ("Bag of Words"). A typical machine learning classification pipeline is shown in Figure 3.

This is extremely challenging in the case of DNA and protein sequences, due to the lack of any "baseline" features, and the differing lengths of whole-protein sequences. We don't have a fixed size (or resizable) digital image with a known number of pixels at each location, or a document with clear "words" or sentences (unlike in natural language).

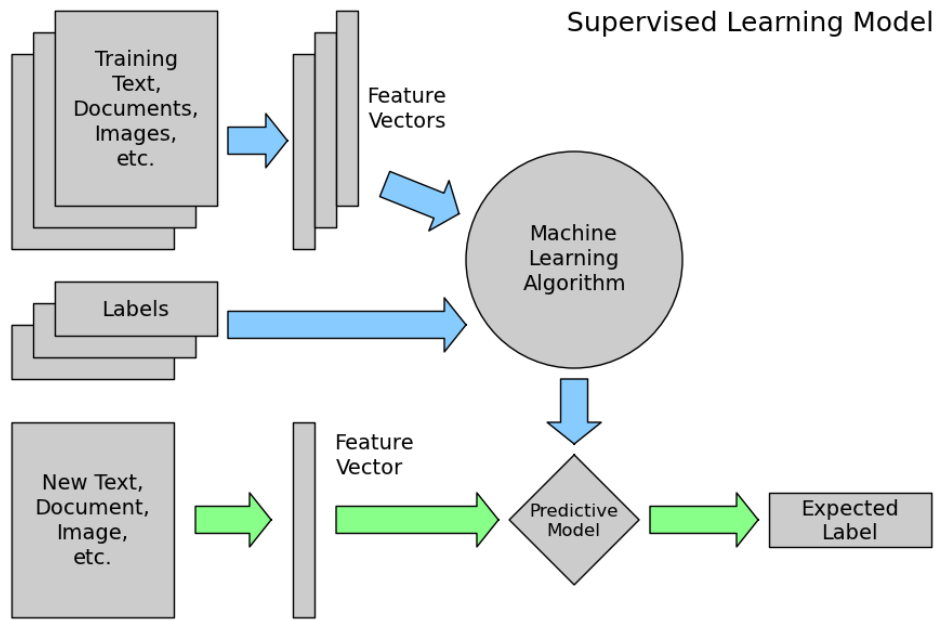

Figure 3 ML classification pipeline. Feature extraction from raw data, ML model-fitting and prediction. Source: SciKit-Learn tutorial (**Pedregosa *et al.*, 2011**)

### 1.2.3 Feature Engineering

"Raw" extracted features are rarely sufficient, i.e. statistically significant, discriminative and distinctive enough. Typically, new features must be created from the baseline feature representation to improve performance. For example, in the case of natural text, instead of counting words, count per document word-frequency (Luhn, 1957). As a more specific example, rather than counting the frequencies of each AA, we can count the frequencies of groups of AA (Reduced AA alphabets) that share a property (e.g. positive charge – Lysine and Arginine). Creating or "engineering" good features is one of the major problems in applied ML (David Kofoed Wind, 2014; Domingos, 2012; Qi *et al.*, 2012; Ofer and Linial, 2015). This problem is currently nigh-impossible to automate generically, and extensive domain-specific knowledge is required to invent novel features for different problems.

### 1.2.4 Feature Selection

Removal of irrelevant and redundant features is important, especially when we can create more features than samples. This can be optional, typically if the number of features is not overly large and the features are non-redundant, uncorrelated and definitively "relevant" (an optimistic scenario at best). At all steps, a potential "feature explosion" (when many possible features may be extracted), (Nanni *et al.*, 2014; Rao *et al.*, 2011; Ofer and Linial, 2015) "the curse of dimensionality" and overall performance must be taken into account .

Feature selection also serves to help us understand the important features in our model, so we can interpret and understand it better. Many different techniques exist for filtering features, notably filter methods (e.g. removal of features according to statistical criteria such as p-values, or highly correlated features), wrapper methods and model-based methods (e.g. deriving variable importance from model weights (Guyon *et al.*, 2002; Guyon, 2003; Peng *et al.*, 2005)).

### 1.2.5   Model Selection

Various learning algorithm exist for data analysis and classification. Various methods have different benefits and drawbacks; for example: prior performance, bias and variance, robustness to noise, suitability for imbalanced classes, training and prediction speed, ease of implementation, interpretability, non-linear feature interaction handling, and many other such considerations. We tested many different models and often ensembled (combined) multiple models (Dietterich, 1990; Karsenty *et al.*, 2014; Ofer and Linial, 2015). Support vector machines (SVM) and decision tree-based classifiers (*e.g.,* Random forests) were nearly always our models of choice, offering excellent performance, interpretability, handling of imbalanced classes and ease of use (Fernández-Delgado *et al.*, 2014; Breiman, 1999; Ben-Hur *et al.*, 2008).

### 1.2.6   Hyperparameter Tuning

Many parameters must be set prior to any model fitting. Such hyperparameters are tuned to optimize model learning capacity (e.g. degrees of freedom) and performance. Such parameters range from the choice of model and preprocessing (e.g. feature selection) to various parameters inherent to each model, such as kernel types in SVMs (Cristianini, 2001; Society), learning rate and network structure in neural networks (Bishop, 2006), the number of trees in RF (Breiman, 1999; Fernández-Delgado *et al.*, 2014; Touw *et al.*, 2013), etc'.

### 1.2.7   Evaluation (Cross-Validation)

Classifier performance is usually evaluated using Cross-validation (CV) (Kohavi, 1995). The model is trained on part of the data, then predictions are made on a separate validation subset which was not seen during model training. Performance is best measured by a models ability to generalize to unknown data.

Data may be further divided into a training set, validation set, and an additional "holdout" test set. A holdout is important when extensive hyperparameter tuning is involved, since the parameters are being adjusted according to the CV performance purely on validation data may be biased, (e.g. feature selection, ANN network architectures).

In *k*-fold cross-validation (Geisser, 1993; Kohavi, 1995), the training set is split randomly into *m* equal disjoint subsets/"folds". In this work, we mostly used stratified K-fold CV (Kohavi, 1995), in which each fold preserves a percentage of samples for each class, making it useful for imbalanced data, avoiding folds populated exclusively by one class. Next, one subset is left out of the training set, and the training is performed on the remaining subsets. Finally, the labels of the instances in the subset which was left out are predicted. This procedure is repeated for each subset, so that "blind" predictions are obtained for the whole training set. The predictions are then evaluated according to some metric such as accuracy.

High performance in cross-validation should indicate that a classifier generalizes well, i.e. will correctly predict new instances as well as it did on known data. However, training data is not always drawn from the same distribution as the real world distribution, nor are its samples necessarily independent and identically distributed (i.i.d.). For example, if "redundant" or near identical sequences are in the training set (e.g. a pair of 99% identical orthologs), then near-identical sequences might be found in both the training and validation set, while in the real world, we would be unlikely to need our model for such "trivial" cases (we'd simply annotate according to sequence similarity). Thus, misleadingly high performance would be obtained by the classifier even if it over-fits the data, since it will have seen a "copy" of the "unknown" sample. This may result in poor generalization and overestimation of model performance. Therefore, it is important to reduce data redundancy (not just between the training and test sets), using tools such as Uclust (Edgar, 2010), CD-HIT (Huang *et al.*, 2010; Sikic and Carugo, 2010) or Uniref clusters (Suzek *et al.*, 2007, 2015). In short, care should be taken to ensure the dataset is non-redundant, and reflects real-world scenarios, rather than an idealized, trivially easy versions. I note that in my experience, many preprocessed "academic" datasets have a marked tendency to be "easier". Competitions such as CASP, CAFA, DREAM and Kaggles' can be a good source of challenging, realistic validation.

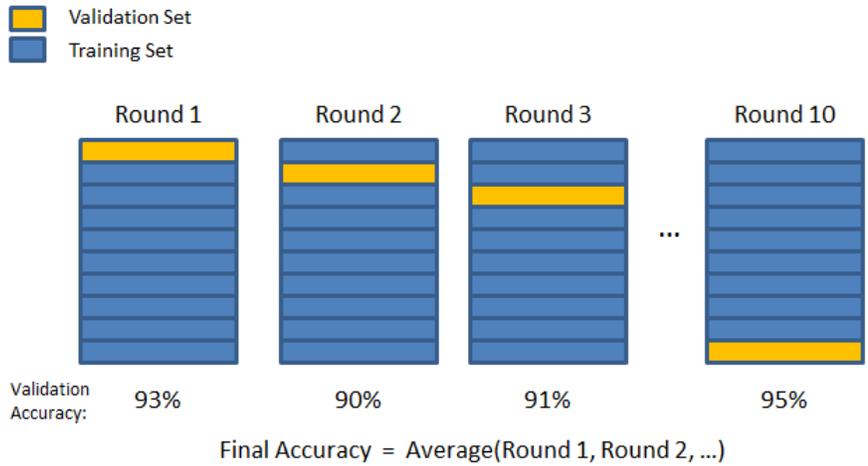

**Figure 4 K-fold (10-fold) Cross Validation illustration. Accuracy can be any metric.**

# 2 מטרות העבודה - GOALS & BACKGROUND

## 2.1 Sequence-alignment methods for protein classification:

Functional annotation is normally done using a supervised approach, *i.e.* inferring functional classification for a sequence according to existing sequences whose functions are known. The problem can be stated as follows: given are an unlabeled protein sequence S and a known protein family (or class) F, We want to determine whether or not S belongs to F. In general, a family is a group of proteins with similar structure and function. If the unlabeled sequence S belongs to F, then we can infer the function of S.

The most naïve, supervised approach is the *nearest-neighbor* search (Sasson *et al.*, 2006), relying on sequence alignment and inferred homology. Given our query sequence, a database of sequences is searched with the goal of identifying similar sequences with preexisting annotations. The most common algorithm for this is BLAST (Altschul *et al.*, 1997), which performs local sequence alignment. If a significantly similar sequence is found, the query sequence will be considered to possess a similar function; i.e. 'guilt by association'. A rule of the thumb for this sort of inference is the '*twilight zone'* concept: a sequence at least 100 amino acids long is likely to be a homolog if at least 30% of its amino acids are identical (Rost, 1999). Below this value the sequence is in the 'twilight zone', where similarity cannot be separated from randomly occurring similarity (using traditional sequence alignment).
Although direct inference is useful for many sequences, it suffers from critical caveats:

(i) In order to learn about a sequence there must exist a significantly similar sequence whose function is known, essentially precluding function prediction for unknown protein families, or those for whom too few "samples" exist to create a statistical model (i.e. a PSSM profile or HMMs (Finn *et al.*, 2014; Remmert *et al.*, 2011; Lavelle and Pearson, 2010)).

(ii) Many proteins with similar sequences have different functions and would therefore be mistakenly classified as having the same function, e.g. *paralogs*.

(iii) Many proteins share functionalities, active sites or domains but possess significantly different sequences, despite having similar functions. E.g. NPPs/Prohormones, wherein the "prohormone" regions flanking the NPs within the sequence are relatively unconserved (Toporik *et al.*, 2014).

(iv) Proteins may share high level function, while sharing little to no structural similarity, let alone sequence based identity. E.g. Disordered proteins (Gitlin *et al.*, 2014; Sharma *et al.*, 2014; Orosz and Ovádi, 2011; Uversky *et al.*, 2000).

(v) Sequence similarity based methods lack significant statistical power given extremely short sequences, i.e peptides. E.g. a pair of sequences of length 200 with 40% similarity might have a pair of shared domains and ~80 aligned AAs. The NPP Pro-opiomelanocortin is just 36 AA long, ~15 AA are rarely a constituent unit of molecular information. Additionally, such short alignments lacks statistical power, i.e BLAST e-value (Akhtar *et al.*, 2012).

(vi) Structural similarity is what *usually* defines function, but proteins with the same structure can have significantly different sequences.

## 2.2  A machine learning approach for protein classification:

When sequence alignment is meaningless, e.g., only a small number of sequences can be aligned, or when the information content of the multiple alignment is minimal, other methods have to be adapted. We approached the problem aiming to develop methods that are not directly sequence-alignment based, but rather rely on extracting alignment-free sequence-derived quantitative features. Each sequence is transformed into a fixed-length vector of features. A discriminative (or generative) statistical model, such as a Support Vector Machine (SVM) (Leslie *et al.*, 2002; Ben-Hur *et al.*, 2008; Cortes and Vapnik, 1995) or decision tree (e.g. RF (Kingsford and Salzberg, 2008; Breiman, 1999, 1996)) is then fitted to the data. Such models can then be used to classify new samples/sequences into the learned classes.

Such methods avoid the requirement for sequence alignments have shown success in learning high-level functional traits (such as the high level of protein family structural folds), while often being far more computationally efficient than NN searches. While both the alignment-based and the sequence-derived feature approaches may use the same information as input, namely the sequence itself, they can perform very differently due to the manner in which they exploit the data and the information they extract from it. There are cases in which an intelligent choice of numerical features (i.e., those that best capture the characteristics of the sequences) can significantly outperform alignment models (e.g., HMM and PSSM), and can be extended for a variety of other problems such as gene co-expression data (Kahanda *et al.*, 2015).

## SUPERVISED LEARNING

In the statistical learning framework of supervised learning, a group of known, annotated samples serve for the learning/training phase. This group of sequences is referred to as the training set. Once the learning stage is complete, the computationally learned hypothesis can be used to classify unidentified samples into a learned class. The advantage of machine learning methods over the sequence-similarity approach (described above) is the fact that the learning methods can take advantage of heterogenous, non-linear features, potentially identifying the minimal conserved set of characteristics in each family of proteins and focusing on searching only for these characteristics. Furthermore, these methods can potentially learn the underlying structure of the problem, discovering unsuspected correlations and important features, and improving as the size of the dataset grows (Halevy *et al.*, 2009). In addition, they may be far faster than the computational demands required for (all to all) sequence-similarity based approaches (Zhao *et al.*, 2012; Edgar, 2004; Hochreiter *et al.*, 2007). These flexible methods can be much more powerful than "brittle", pre-defined rules (e.g. regular expressions) or the naïve sequence similarity approach.

The field of machine learning is immense with a strong impact on biology (Ben-Hur *et al.*, 2008; Tarca *et al.*, 2007). Machine learning algorithms include SVMs and kernels (Lodhi, 2012; Leslie *et al.*, 2004; Cristianini, 2001; Kuang *et al.*, 2005), decision trees, Neural Networks (NNs) (e.g. Extreme learning machines (Huang *et al.*, 2015; McDonnell *et al.*, 2014), Convolutional NNs, Recurrent NNs (Sønderby *et al.*, 2015; Lipton, 2015; Hochreiter *et al.*, 2007)), decision trees, ensembles of classifiers (such as Random Forests and AdaBoost (Breiman, 1999; An Empirical Comparison of Supervised Ensemble Learning Approaches)), and unsupervised methods for clustering (Rappoport *et al.*, 2010).

## 2.3 Defining levels of Protein "Function"

Protein "function" can be defined at many levels – both in terms of molecular function, biological processes and cellular localized activity (in the manner of GO (Ashburner *et al.*, 2000)), and in terms of "high/low level" function. For example, whether a protein is a membrane bound receptor, a G-protein coupled receptor (GPCR), a GPCR that uses cyclic AMP (Alberts *et al.*, 2009), or an olfactory GPCR sensitive to CO2 (Tauxe *et al.*, 2013; Lu *et al.*, 2007; DeGennaro *et al.*, 2013; Hallem and Carlson, 2006).

There is an obvious connection between the 'granularity' of a function and the evolutionary diversity of the proteins that share it (Rost, 1999; Kobayashi *et al.*, 1998; Dubchak *et al.*, 1995). Typically, groups of proteins that share a high-level functionality (*i.e.,* enzymes or membrane receptor) are larger and more diverse than low-level (e.g., urease enzymes) functional groups (Shachar and Linial, 2004). Community competitions in functional assignment, such as the Critical Automatic Functional Annotation initiative show that there is considerable room for improvement (Radivojac *et al.*, 2013; Gillis and Pavlidis, 2013; Bacardit *et al.*, 2014).

While inferring precise, "low level" functionality is often of interest (e.g. annotation of functional homologs (Loewenstein *et al.*, 2009; Bork and Koonin, 1998)), there is an obvious interest in successfully learning *high-level functionality* (Sasson *et al.*, 2006; Naamati *et al.*, 2009). If the training set consists of proteins that share a low-level functionality, the classifier would only be able to detect proteins that belong to the narrow function that was learned. However, if the training set consists of proteins that share a high-level functionality, the classifier will be able to detect any protein that belongs to a very broad class, even if 'close' representatives are unknown.

As an example, consider the case of the Major Facilitator Superfamily (MFS). This superfamily includes over 300,000 proteins capable of transporting small solutes in response to ion gradients. In general terms, proteins of the family belong to the transmembrane transport system. However, if we use classifiers of low-level functionality, we would have a classifier for the different types of MSF families (including Nitrate transporter, Sialic acid transporter and many more). If a gene for a novel subtype of MSF were found, we would not be able to identify its function at the lower level as it would not belong to any known transporter family. However, if our classifier was trained to identify "MSFs", we could identify the sequence as a (novel) MSF.

It is difficult to learn high-level functionality, particularly when we might barely understand them on the theoretical level. While we might expect nitrate transporters from different organisms to share similar sequences due to evolutionary homology, we would not expect this of a high-level group such as all transmembrane transporters.

Structure (rather than sequence) is the major element for most (but not all!) proteins' function (Orosz and Ovádi, 2011; Berg *et al.*, 2002). Structural knowledge and predictive capabilities is severely lacking, despite improvements in computational prediction thanks to the ongoing, targeted mapping of the structural genomic-fold space (Portugaly and Linial, 2000), and improvements in predictive methods (Magnan and Baldi, 2014; Zhou and Troyanskaya, 2014; Sønderby and Winther, 2014; Remmert *et al.*, 2011). Experimental determination of structure using NMR or X-ray crystallography remains slow, expensive and impossible for many proteins (Carpenter *et al.*, 2008), with just ~110,000 determined structures for tens of millions of proteins (Berman *et al.*, 2000; Rappoport *et al.*, 2012; Boutet *et al.*, 2007).

## 2.4 Short proteins & peptides: an overlooked niche

The ability to learn about a protein by comparing it to its (inferred) homologs has been used in functional prediction, secondary structure prediction, three-dimensional fold prediction and several other applications (Qi *et al.*, 2012). However, the power of sequence alignment/similarity-based tools is greatly diminished for short proteins. This is because when comparing short sequences it is difficult to distinguish genuine homology from mere evolutionary noise/coincidence. For example, submitting a short amino acid sequence to a sequence-similarity search server such as BLAST will usually result in matches with insignificant e-values (a statistical measure of significance for the expectation value), even for sequences with high percentages of identity. Therefore, the detection of short proteins' homologs using sequence-alignment tends to fail (Naamati *et al.*, 2009).

The difficulty in the identification of homologs is only one of the problems associated with short proteins (Frith *et al.*, 2006). Consider a newly sequenced genome. The first task is identifying potential (putative) gene products. The main steps for identifying encoded proteins include:

- o Sequence similarity: While this method is the most powerful computational approach, as indicated, it fails to detect short proteins.

- Comparative genomics: This method requires the aligned genomes of related species. Additionally, this method is likely to fail to detect short proteins for similar reasons to the sequence similarity approach.
- Ab-initio gene prediction: The default parameters require a minimal length for potential ORFs (Open Reading Frames), which may further hinder the detection of short proteins.
- High coverage of the transcriptome and proteome by high throughput (HT) technologies.

Some experimental methods focus on detecting mRNA expression and others on protein expression. HT methods such as RNA-Seq (Wang *et al.*, 2009), Ribosomal profiling (Ingolia *et al.*, 2012) and Mass Spectrometry (MS) (Aebersold and Mann, 2003) are perhaps the best source of experimental data for detecting new proteins, but are often far from comprehensive due to the fact that many genes are only expressed under certain conditions (Ponting and Grant Belgard, 2010). Furthermore, discriminating between peptides and "noise" (e.g. random fragments) is extremely challenging, with even the best high-throughput protein expression methods such as tandem MS (MS/MS (Angel *et al.*, 2012)) requiring special tweaking and is limited to the detection of known, highly expressed proteins (Akhtar *et al.*, 2012; Salisbury *et al.*, 2013; Vogel and Marcotte, 2008; Kim *et al.*, 2011; Craft *et al.*, 2013). Often, if a short protein is not already a known candidate, it will not be found (Lubec and Afjehi-Sadat, 2007). As a consequence of these computational and experimental difficulties, short proteins represent a relatively understudied and neglected niche (Tirosh *et al.*, 2013; Frith *et al.*, 2006; Tirosh *et al.*, 2012; Naamati *et al.*, 2009; Su *et al.*, 2013).

### 1.2.4 Neuropeptides: Behavioral & physiological regulators

Neuropeptides (NPs) are short proteins (peptides) that are produced and secreted from nervous system cells (e.g. Neurons), acting as inter-cellular modulators and messengers. They are known to be key modulators in behavior, sensation and homeostasis. They function in neurobiological communication for all Metazoans, with similar roles in Cnidarians and Bilaterians (Mirabeau and Joly, 2013; Semmens *et al.*, 2015).

The NPs are very short active peptides (~5-30 amino acids) produced from longer precursor proteins (NPPs) following proteolytic cleavages. The post-translational end products are subsequently modified and secreted as the actual NPs.

NPs normally locally modulate presynaptic or postsynaptic cell activity (Root *et al.*, 2011;

Nathoo *et al.*, 2001; García-López *et al.*, 2002; Brain and Cox, 2006). NPs typically work by binding cell surface GPCRs, resulting in the initiation of a signaling cascade (Tanaka *et al.*, 2014; García-López *et al.*, 2002; Hewes and Taghert, 2001; Root *et al.*, 2011).

From a functional perspective, known effects of NPs include stress control (Chang and Hsu, 2004; Hökfelt *et al.*, 2000), pain perception, social behaviors and sleep-wake cycle (Nässel, 2002) , food uptake (Nguyen *et al.*, 2011; Taghert and Nitabach, 2012), selective appetite (Beshel and Zhong, 2013; Dillen *et al.*, 2013) and more. "Social" NPs such as oxytocin (OXT) and arginine vasopressin (AVP) regulate mating (e.g. PBAN - activates pheromone biosynthesis (Rafaeli, 2009)) , complex social cognition and behavior (including pair-bonding, social recognition and maternal behavior) (Gruber and Muttenthaler, 2012; Ferguson *et al.*, 2000; Insel and Young, 2000).

NP sequences are mostly non-alignable due to their shortness, diversity, and lack of sequence similarity (Clynen *et al.*, 2010). Sequence similarity methods fail to predict or provide a comprehensive catalogue of NP bioactive peptides or their precursors. Their immense diversity of sequence, targets, pattern of post-translational modification and specificity (García-López *et al.*, 2002; Hewes and Taghert, 2001) reflects their equally immense range of behavioral and physiological effects in diverse organisms (Nathoo *et al.*, 2001; García-López *et al.*, 2002; Shaw, 1996; Brockmann *et al.*, 2009).

1.2.5  Neuropeptide Processing

The majority of NPs are produced from a larger precursor (NPP, as a result of a series of post-translational modifications and cleavages  (Veenstra, 2000). NPP can produce multiple copies of different NPs as well as multiple copies of each NP (Smit *et al.*, 1992; Fisher *et al.*, 1988; Nässel, 2002). Notably, a cluster of basic residues specifies these cleavage sites by the family of Prohormone Convertases (PCs), intracellular endopeptidases such as Furin (Veenstra, 2000). However, many NPs  (*e.g.,* cathepsin L (Funkelstein *et al.*, 2010)) do not obey the dibasic residues specificity rule. Furthermore, only a fraction of sites containing the "canonical" ("Known Motif" (Southey, Rodriguez-Zas, *et al.*, 2006)) dibasic cleavage pattern are actually cleaved (Devi, 1991; Southey, Rodriguez-Zas, *et al.*, 2006; Amare *et al.*, 2006; Southey, Amare, *et al.*, 2006; Southey *et al.*, 2008; Tegge *et al.*, 2008; Amare and Sweedler, 2007). Many non-secreted, non-NPPs proteins also undergo this cleavage!

### 1.2.6 Neuropeptide Discovery

One goal of this research was to enable the systematic identification of NPPs (and NPs) at a genome-wide scale. The difficulty in identifying NPs and classifying genes as potential NPs stems from the following: (i) Current gene annotation tools mostly rely on sequence conservation traits (Loewenstein *et al.*, 2009) . However, NPs exhibiting the same function may share minimal sequence similarity, and can even lack 3-D structural similarity (Clynen *et al.*, 2010). The NPPs too can vary immensely, sharing only varying amounts of different internal NPs. Additionally, homologous NPPs may still produce NPs that are species specific.

 (ii) Structural inference tools (Lobley *et al.*, 2009; Radivojac *et al.*, 2013; Kuznetsov and Rackovsky, 2003; Midic *et al.*, 2009) fail when applied to short peptides.

 (iii) NPPs undergo distinctive cleavage by PCs, but they're not unique in this regard! Thus, even the presence of identified PC cleavage sites, via tools such as NeuroPred (Southey, Amare, *et al.*, 2006; Southey *et al.*, 2008) or our own forthcoming ASAP/CleavePred[1], is not enough to identify NPPs. Rather, these can only help identify NPs given a NPP.

Consequently, assigning functions to known NPPs and identifying previously overlooked genes calls for developing an alternative strategy. Methods based on sequence alignment don't work, and strict, high-bias (Hastie *et al.*, 2009), "hand-designed" rules (e.g. regular expressions or motifs (Dinkel *et al.*, 2014)) lack the capacity to model this problem. Towards this goal, we developed our ML approach.

---

[1] Forthcoming work. Code and methods available at: https://github.com/ddofer/asap

## 2.5 Goals

To summarize, my research had a number of goals which were answered through this work. These are collated in a number of peer-reviewed articles (attached in the appendix). The goals and articles include:

- o A universal feature extraction & ML framework for proteins, suited for classification of high-level functions and unified properties, without external databases or alignment.
- o Developing a universal set of feature extraction methods and novel engineered features for efficient feature representations of whole protein sequences.
- Implemented as ProFET (Protein Feature Engineering Toolkit) (Ofer and Linial, 2015)[2].
- o Creation of a publicly available platform for identifying novel Neuropeptides precursors & Neuropeptides on a massive scale.
- Implemented as NeuroPID (Ofer D *et al.*, 2014)[3], and the NeuroPID web-server (Karsenty *et al.*, 2014): neuropid.cs.huji.ac.il
- o Application of the Neuropeptide classifier to unknown proteins and genomes to discover putative Neuropeptides.
- Performed as part of NeuroPID (Ofer D *et al.*, 2014), on multiple organisms, as well as further work[4]; See Appendix – a sample of predicted Cnidarian sequences. (Unpublished).

---

[2] https://github.com/ddofer/ProFET

[3] NeuroPID source code freely available at: http://www.protonet.cs.huji.ac.il/neuropid

[4] www.protonet.cs.huji.ac.il/neuropid/results

# 3 שיטות עבודה — METHODS

## 3.1 Protein datasets

In gathering the protein datasets in this study, we used (i) custom sets gathered from public databases such as UniProtKB/UniRef (Suzek *et al.*, 2007; Boutet *et al.*, 2007) and SCOPe (V2.05) (Fox *et al.*, 2014) and (ii) benchmarks extracted from publications. For both resources, we applied CD-Hit (Huang *et al.*, 2010) and USearch (Edgar, 2010) to remove redundant and similar sequences according to a predefined % threshold of sequence identity. As a rule, we used only classes that contained a minimal number (~40) of samples per group (after redundancy removal). Sequences with unknown amino acid (AA), errors or sequences that are shorter than 30 AA were removed. I note that different versions of the NPP dataset were used: V1 was used in the original NeuroPID paper. V2 was used in ProFET and was based on V1, with various changes, namely a different negative set.

The datasets are:

### Specialized functions

- Ribosomal proteins: Acquired from SWP and partitioned to Archea, Bacteria and Eukarya. Redundancy filter was set to 20–40% identity (according to the set size).
- Thermophilic proteins: The ThermoPred (Lin and Chen, 2011) dataset was used, consisting of 915 thermophilic and 793 non-thermophilic (Mesophile) proteins. Further filtered for < 40% sequence identity.
- DNA-binding proteins. Benchmark dataset from DNA binder (Kumar *et al.*, 2007).
- RNA-binding proteins. Benchmark dataset from BindN (Wang *et al.*, 2010).

### Subcellular localization:

- LocTree3 benchmark (Goldberg *et al.*, 2014) for Eukarya and Bacteria were used. Filtered at 40% identity within each class.

- Mammalian subcellular localization: Protein-organelle pairs acquired from SWP.
- Uncultured bacterium sequences extracted from UniProtKB and mapped by keyword annotations for major cellular compartments (membrane, cytoplasm, ribosome). Filtered by UniRef50 clusters. 15,995 sequences.

### Structural classifications:

- SCOPe (Release 2.05, Feb 2015) (Andreeva *et al.*, 2014; Fox *et al.*, 2014). Classes and folds were defined by SCOP, with 25% or 10% sequence identity filter. (8,514 and 6,721 sequences respectively).
- SCOPe (Release 2.05, Feb 2015) "partial classes" used - defined by the SCOP class (marked a-k), with classes c,d removed. We also apply as a benchmark classes "a,b,f,g" at 25% sequence identity filter. Classes a,b,c,f,g were tested following similarity removal at an extremely low level (10%).

### Viral properties and classes:

- Virus-Host pairs: Acquired from SWP. This set includes all viral proteins partitioned by the kingdom of the hosts. Redundancy filtration (at 40% identity) was performed on the viral proteins but not on the hosts.
- Capsids: Compilation of two sets of all viral capsid proteins annotated by SWP: (i) Classes according to host type. (ii) Classes according to viral replication mode.

### Neuropeptides

- Neuropeptide precursors (NPPs): Proteins annotated as 'neuropeptide' were acquired from SwissProt (SWP) and UniRef90 as a positive set. We removed proteins containing the terms 'fragment' and 'receptor'.
    a. In the original NeuroPID dataset, multiple non-overlapping negative sets were constructed by randomly sampling proteins with the same distribution of lengths. Multiple additional sets were constructed seperately, see results, & original NeuroPID article (Ofer D *et al.*, 2014) for details.
    b. In the V2 (ProFET) NPP dataset: we used as our negative set proteins containing Signal peptides and lacking a validated TMD. We keep all such sequences with the

same (atypical) range of lengths as the NPPs. The negatives were filtered for 10% identity. The dataset held 2,309 negatives and 1,269 NPPs. Note that we expect many unidentified NPP peptides amongst the "negatives".

## 3.2 Features

All ProFET features are extracted directly from the protein sequence, and do not require external input (Saeys *et al.*, 2007). NeuroPID used a subset of the features implemented in ProFET. ProFET and NeuroPID were implemented using Python (Rossum and Drake, 2010). Users can easily add additional features. ProFET also generates a predefined set of default features for consistency in evaluation and ease of use, callable from the command-line.

Many of the features can be restricted to a segment of a protein (e.g., the first third of a sequence). The activation of global features combined with segmental consideration is motivated by the atypical composition of different segments in numerous proteins, e.g. signal peptides (Petersen *et al.*, 2011), flexible N-terminal linker regions (George and Heringa, 2002), transmembrane domains (Käll *et al.*, 2007), disordered regions (Linding *et al.*, 2003; Deng *et al.*, 2015) and more. Sequence similarity, as measured by PSI-BLAST was extracted for some experimental comparisons, but was not implemented in our framework.

The features extractable by ProFET include:

### Biophysical properties:

Many of these properties were derived from the Expasy proteomics collection (Tools *et al.*, 2010). The important of these elementary global features has been previously validated (Varshavsky *et al.*, 2007). All of these features are included in the NeuroPID framework.

 i. Molecular weight (in Da).
ii. Sequence length (in AA).
iii. pH (I), the isoelectric point (Hobohm and Sander, 1995).
iv. Net Charge at various pH (I)s.
v. Aromaticity, the relative frequency of Phe, Trp, Tyr.
vi. Instability index, an estimate for in-vitro protein stability (Wilkins and Williams, 1997).
vii. GRAVY (Grand Average of Hydropathy), the sum of hydropathy values of all AA, divided by the number of AA in the sequence (Kyte and Doolittle, 1982).

viii. Aliphatic index, the relative volume occupied by aliphatic side chains (Ala, Val, Ile and Leu) (Ikai, 1980; Wilkins and Williams, 1997).

Letter-based features:

i. AA compositional frequency. i.e we counted each amino acid for each sequence, then divide counts by length. This is equivalent to Term-Frequency (TF) features, or a K-mer of size K=1 (see below).

ii. Overlapping K-mers (Kuang *et al.*, 2005; Leslie *et al.*, 2002). K-mers are combinations of K letters, counted in overlapping windows within the sequence.

iii. "Mirror" K-mers. This accounts for symmetrically equivalent K-mers. For example, Lysine-Arginine (KR) is counted together with RK.

iv. Reduced amino acid alphabets. Grouping AAs according to shared physicochemical properties secures compact representations. We used handpicked reduced alphabets from the literature (Murphy *et al.*, 2000; Etchebest *et al.*, 2007; Weathers *et al.*, 2004; Peterson *et al.*, 2009; Susko and Roger, 2007) and novel alphabet representations of our own (e.g. Ofer_14 and Ofer_8). For the 14 AA alphabet the following are grouped together: KR, TS and LIVM. For the 8 AA representation the grouping is for FYW, ALIVM, RKH, DE and STNQ. The other AA remain uncompressed. This was combined with the various K-mer and the various segmental composition features. E.g. the frequency of 3 letter alphabet 4-mers, grouped as {K,R}, {P}, {The remaining letters}, (3^4 = 81 features).

Local potential features:

i. Potential post-translational modification (PTM) sites. We included short motifs implemented as regular expressions, including those for "known short motif " dibasic cleavage model (X-X-Lys-[Lys or Arg], X-X-Arg-Arg, Arg-X-X-[Lys or Arg]; where X denotes any AA (Southey, et al., 2006; Veenstra, 2000). Others include potential N-glycosylation, Asp or Asn hydroxylation motifs and a Cys spacer motif that captures the tendency of Cys to appear in a minimal window (Naamati *et al.*, 2009).

ii. Potential Disorder (FoldIndex). Local disorder is predicted using the naive FoldIndex (Prilusky *et al.*, 2005) and TOP-IDP methods (Campen *et al.*, 2008). FoldIndex predicts disorder as a function of hydrophobic potential and net charge.

## Information based statistics:

These features aim to capture the non-random distribution of AA in the sequence, based on the concept of information entropy. For details, see section 2.3, (C) in the original paper (Ofer D *et al.*, 2014).

The information-based features used are:

i. Amino acid entropy. This set of features captures a property how non-randomly distributed each amino acid is in the sequence, based on the concept of molecular information entropy. (Schneider, 2010; Ofer D *et al.*, 2014)

ii. Autocorrelation. This

ii. Binary autocorrelation & entropy. Selected letter (s), for example, K, R and C are denoted as "1" and the rest as "0". Lag and entropy are then computed for this new representation.

## Amino acid propensity-scales:

Amino acid (AA) propensity scales map each AA to a quantitative value that represents physicochemical or biochemical properties, such as hydropathicity, secondary structure propensity (Jr and Fasman, 1989) or size. Thus, dissimilar AA (for a given property or statistical propensity) will have more dissimilar values. These scales can then be used to represent the protein sequence as a time series, typically using sliding windows of different sizes. ProFET includes a wide array of scales, ranging from the established propensities for hydrophobicity and flexibility/B-factors (acquired from Expasy (Tools *et al.*, 2010)), to independent derived scales (Atchley *et al.*, 2005; Georgiev, 2009).

Features derived from these scales include:

i. Averages for the sequence as a whole, for different window sizes or segments.
ii. Quartile averages (*e.g.,* average of the top, bottom 25%).
iii. Maximum and minimum values for a given scale and window-size along the entire sequence.
iv. Autocorrelation.

## Transformed CTD features:

We implemented multiple CTD (Composition, Transition, Distribution) features, some based on Dubchak (Dubchak *et al.*, 1995) and PROFEAT (Li *et al.*, 2006) as well as novel ones based on AA propensity scale ranges and reduced alphabets. These 3 letter groupings included:

hydrophobicity, normalized Van der Waals volume, polarity, polarizability, charge, secondary structure and solvent accessibility hydrophobicity, normalized Van der Waals volume, polarity, polarizability, charge, secondary structure and solvent accessibility propensity. An additional subdivision of disorder propensity was adapted from Composition Profiler (Vacic *et al.*, 2007): 1:'ARSQEGKP', 2:'ILNCFYVW', 3:'DHMT'.

These features are:

i. Composition (C) is the number of AA in a grouping divided by the total number of AA.

ii. Transition (T) is the number of transitions from a particular property to different property, divided by (total number of AA − 1).

iii. Distribution (D) captures is the chain length within which the first 25%, 50%, 75% and 100% amino acids of a particular property are located.

## 3.3 Evaluation

As mentioned previously, for each dataset performance was measured using multiple rounds of stratified cross validation or stratified shuffle-split CV (a bootstrapping with replacement variant). Multiple metrics were used and recorded, with an emphasis on metrics that are robust in the case of imbalanced classes, e.g. F1, ROC-AUC (area under the Receiving Operator Characteristic curve), see Figure 5.

Hyperparameters were tuned solely according to the training data's CV performance, without exposure to the test set. Feature selection (where done) was performed independently by each training fold. For the exact details on the chosen models and parameters, I refer readers to the original articles (Ofer and Linial, 2015; Ofer D *et al.*, 2014).

$$\text{Accuracy} = \frac{TP + TN}{TP + FP + TN + FN}$$

$$\text{Precision} = \frac{TP}{TP + FP}$$

$$\text{Recall (sensitivity)} = \frac{TP}{TP + FN}$$

Figure 5 Metrics

## 3.4 Machine Learning Methods

A large variety of machine learning methods were tested. The best performing methods included SVMs (with RBF kernels), Random Forests, Gradient boosting trees, Logistic Regression and ensembles of different models. All were implemented with Scikit-Learn (Pedregosa *et al.*, 2011).

## 3.5 Feature selection (RF-RFECV)

Interpretability of the features that best contributed to the performance is a crucial knowledge. Several methods for feature selection can be applied to identify a minimal set of such features, or to remove "noisy" features. In addition to classic statistical filters (e.g. FDR corrected F-test), we applied a novel method of my own devising I named RF-RFECV. It consists of combining Random Forests (an ensemble of decision tree classifiers, where each component has only a subset of the features) with the Random Feature Elimination (RFE) wrapper method.
In each iterations, the weakest features are removed, and the model is then retrained with the remaining features, until the preselected desired amount of features remains. Performance of the reduced feature set is measured using new splits of the training data and cross validation. This is then repeated. The existing RFE implementations typically used an SVM or linear learners and not ensembles or random subspace based learners.
This method was applied for "aggressive" post-hoc feature selection, with the goal of extracting interpretable features from the hundreds of weak, possibly redundant features. This method combined the powerful modelling capacity of RF with the RFE method, giving a tiny subset of powerful, non-linear features within a short time. We applied it to the various datasets with excellent results(Ofer and Linial, 2015).

## 3.6 Software and implementation

The framework was implemented using Python (Rossum and Drake, 2010). Sci-Kit Learn (Pedregosa *et al.*, 2011) was used for ML and statistical analysis. Additional libraries included BioPython, NumPy, SciPy and Pandas (Cock *et al.*, 2009; Oliphant, 2007; McKinney, 2011).
The code is freely available online: https://**github**.com/ddofer/**ProFET** ,
http://www.protonet.cs.huji.ac.il/neuropid/

# 4 תוצאות Results

## 4.1 Platforms

We developed Neuropeptide Precursor Identifier (NeuroPID), a platform to automatically identify neuropeptide precursors in metazoan genomes, and through them, their secreted bioactive peptide products. The NeuroPID platform was successfully applied to multiple genomes, including insects, Cnidarians (Appendix) and Octopus Vulgaris (Ofer D *et al.*, 2014), identifying hundreds of high confidence candidates. Furthermore, we expanded the tool into a freely available, easy to use online webserver, capable of easily handling tens of thousands of sequences within minutes and annotating whole genomes.

ProFET (Protein Feature Engineering Toolkit) expanded our systematic approach into a comprehensive platform for all possible proteins. It builds on NeuroPID's underlying framework with new methods and feature extraction techniques, as well as a comprehensive set of routines for managing the data (e.g. raw sequences, annotations, extracted features and predictions). It ranks among the state of the art for rapidly classifying high-level novel functionality in proteins, without external databases or alignment.

ProFET serves as a platform for machine learning and quantitative sequence-derived feature extraction from proteins. This did not exist before, with various scientists constantly reinventing the wheel, rewriting "glue code", and only looking at "one-off" cases, rather than viewing this as a paradigm. In addition to utility, we provided performance, with a wealth of universal, powerful features that can be extracted from any sequence, without needing prior annotation, homologs or structural knowledge. This is notable, as a large number of existing tools are unsuited for large scale predictions or work with novel proteins due to their requirements (Saraç *et al.*, 2010). We thus created a universal methodology for machine learning with whole protein sequences.

## 4.2 NeuroPID Results

The performance of the NeuroPID platform and datasets is presented here in brief, in Figure 6 and

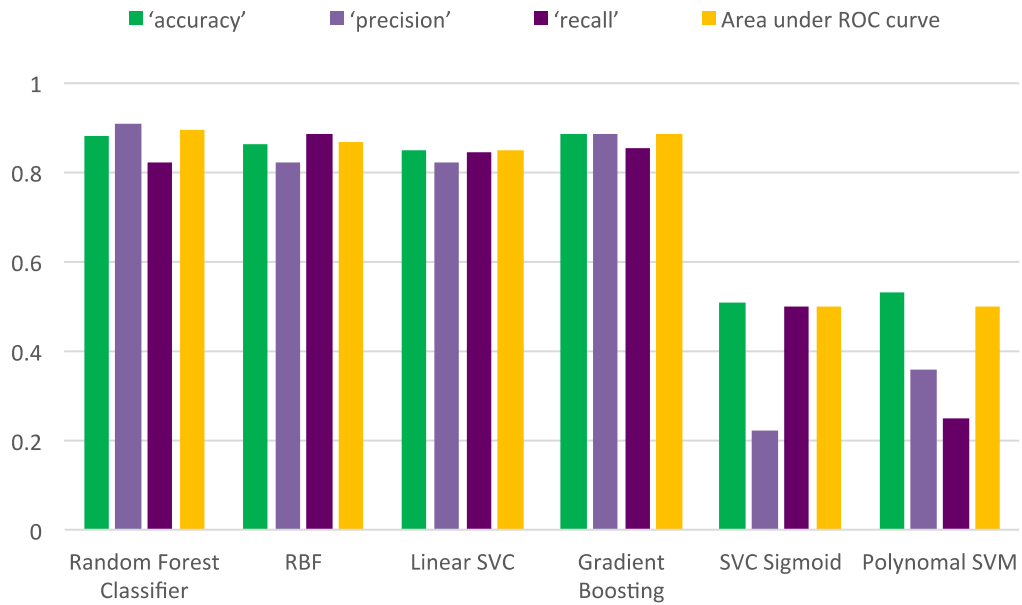

Table 1. For full details, I refer readers to the original article, included in the appendix.

Figure 6 Cross validation performance of different models on NeuroPID's full, original dataset (NPP V1). Numerous models support the high accuracy and precision in the CV tests. Result variations measured by repeated tests were negligible. RBF = SVM with Radial Basis Function Kernel. SVC = SVM

| ML model\Metric | Accuracy | Precision | Recall | MCC | AUC |
|---|---|---|---|---|---|
| Random Forest | 0.928 | 0.938 | 0.915 | 0.857 | 0.928 |
| Gradient Boosting | 0.93 | 0.928 | 0.929 | 0.859 | 0.93 |
| Linear SVM | 0.881 | 0.87 | 0.894 | 0.763 | 0.882 |
| SVM-RBF | 0.928 | 0.918 | 0.939 | 0.857 | 0.929 |

Table 1. Performance of the top ML classification models implemented in NeuroPID. Webserver predictions use the first three models.

### 1.2.1 NeuroPID: Alternate Negative Set Performance

To assess the robustness of NeuroPID we repeated the CV procedure with models trained on alternative "negative" sets (the positive set remained unchanged). We performed it for "negative" sets of secretory proteome (TMD containing proteins) and charged nuclear proteins. The rationale behind the selection of the later set was to provide a "challenging" set in which the inherent appearance of basic residues prevails (*e.g.,* histones, transcription factors). CV performance was very high, even exceeding the original results. We did not use these alternative datasets in the final platform, for reasons that will be discussed later.

| Metric \ Negative Set | Nuclear proteins | + SP +TMD |
|---|---|---|
| Accuracy | 0.882 | 0.967 |
| Precision | 0.884 | 0.916 |
| Recall | 0.841 | 0.89 |
| Area under ROC curve | 0.878 | 0.937 |

Table 2 NeuroPID Alternative negative sets performance. Performance is for a Random Forest classifier and original positive training set. SP+TMD = Swissprot proteins annotated as having a transmembrane domain and a signal peptide. Both sets were filtered for redundancy.

### 1.2.2 NeuroPID: Taxonomical Evaluation

Once the ML was trained and tested, we examined unseen examples derived from complete proteomes. We defined the task as identifying unseen NPPs from specific Taxa (Chordata and Arthropods). Only proteins belonging to the taxon were used to construct the negative and positive data. Only representative sequences from UniRef90 clusters were used. The results are summarized in **Error! Reference source not found.**. The high performance of taxon specific models shows that restricting predictions and data to specific taxa typically increases performance (compared to

| ML model\Metric | Accuracy | Precision | Recall | MCC | AUC |
|---|---|---|---|---|---|
| Random Forest | 0.928 | 0.938 | 0.915 | 0.857 | 0.928 |
| Gradient Boosting | 0.93 | 0.928 | 0.929 | 0.859 | 0.93 |
| Linear SVM | 0.881 | 0.87 | 0.894 | 0.763 | 0.882 |
| SVM-RBF | 0.928 | 0.918 | 0.939 | 0.857 | 0.929 |

Table 1), i.e. is easier, despite the reduction in data samples.

|  | Chordata | Arthropods |
|---|---|---|
|  | SVM/Gradient Boosting | SVM/Gradient Boosting |
| Mean Accuracy | 0.746 / 0.907 | 0.900 / 0.923 |
| Mean Precision | 0.906 / 0.923 | 0.968 / 0.941 |
| Mean Recall | 0.621 / 0.914 | 0.889 / 0.952 |
| Mean AUC | 0.768 / 0.906 | 0.908 / 0.901 |

Table 3 NP classification CV with taxonomical partitions. SVM = Support Vector Machine with RBF kernel. Gradient Boosting = Ensemble of Gradient Boosting Trees.

### 1.2.3 NeuroPeptide Statistics

As part of our analysis, we did an exploratory analysis of the biophysical features' distributions in the set of known Neuropeptides vs the background distribution, consisting of all (UniRef50 clustered) proteins with roughly the same range of lengths (under 900 AA long) as the NP. These results confirmed the atypicality of NPs properties, and the validity of our basic features prior to further work (i.e adding more features or model fitting). Results appear in Appendix 3.

We performed a similar analysis on the final set of training data (the final positive set, and the randomly sampled negative set), some significant features are listed in statistically significant different values (Two sided t-test, α=0.01; KS test< 0.01). The most individually significant feature is the occurrence of lysine (K, p-value <1.0e-83).

Figure 6.Figure 7 Features' statistical significance for NPs vs the background (UniRef50) distribution for: (A) amino acid frequency. (B) Several biophysical properties. Most features had

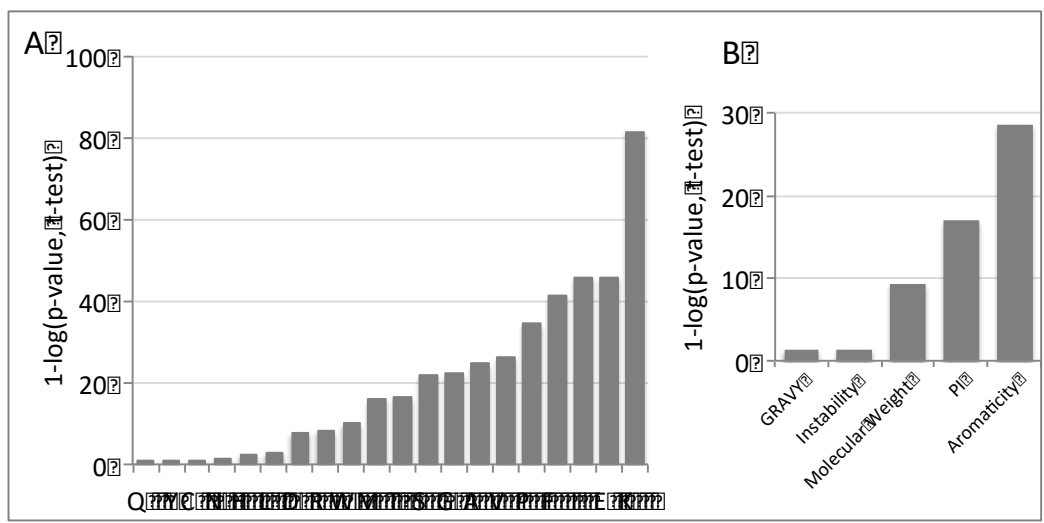

Figure 7 Features' statistical significance for NPs vs the background (UniRef50) distribution for: (A) amino acid frequency. (B) Several biophysical properties. Relative importance shown. P-values obtained from a 2 sided t-test.

### 1.2.4 NeuroPID Webserver

As part of our efforts to ensure the use of our work by the wider scientific community, we created an online webserver for NP and NPP discovery. The NeuroPID website is available at http://neuropid.cs.huji.ac.il.

For full details, I refer to the article (Karsenty *et al.*, 2014), included in the Appendices.

In brief, the webtool can handle genomic scale sequence inputs, and outputs predicted Neuropeptide sequences. Analysis of the features, likelihood of the results, and filtering (e.g. "Display only high confidence predictions, from input sequences with a predicted signal peptide") is also provided. Links to other websites are provided for further analysis (e.g. NeuroPred (Southey, Amare, *et al.*, 2006; Southey *et al.*, 2008), for predicting cleaved NPs). An illustration of some features provided by the website is presented in Figure 8**Error! Reference source not found.** :

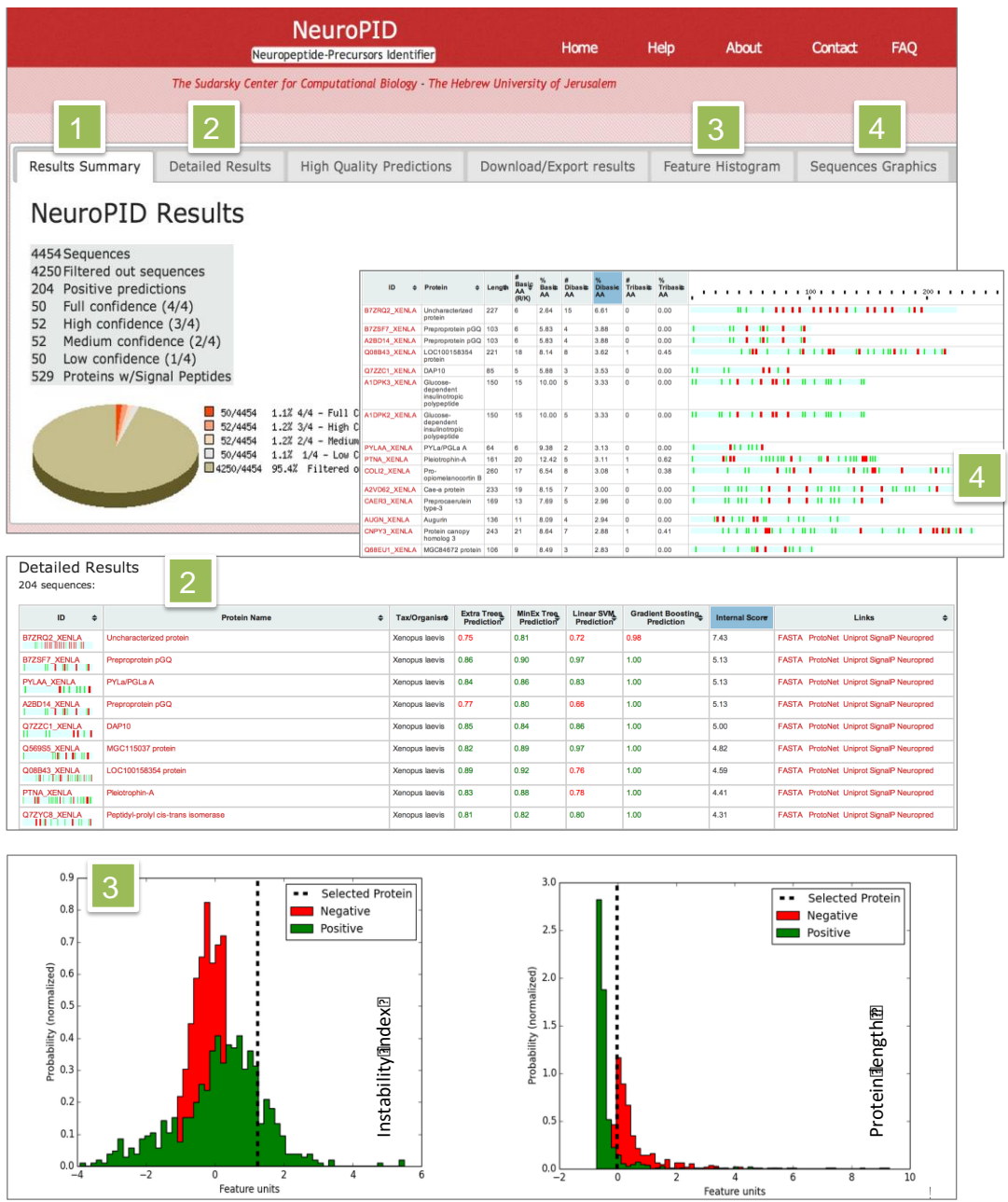

Figure 8 Screenshots of the NeuroPID Website (1) A summary table for an input of 4454 proteins from Xenopus laevis. A pie chart displays the distribution of prediction according to the agreement of the 4 classifiers. (2) A detailed table shows the confidence for each prediction methods. The red and green fonts indicate negative and positive predictions, respectively. The table is ranked by the Internal Score (IS). Results are linked to their FASTA sequence and to knowledge based resources (ProtoNet, UniProtKB) and analysis tools (SignalP, NeuroPred). (3) Results from the feature histogram. The position of a selected protein (dash line) in view of the distribution of the positive and negative sets is shown. The basis distributions are shown in green and red colors to indicate positive and negative instances, respectively. Data is normalized by the distributions standard deviations (denoted Feature Units, X-axis). (4) Sequence graphics with positions of basic residues. The red vertical lines show the location of dibasic and tribasic residues. The presence of single basic residues is colored green.

### 1.2.5 Neuropeptide Taxonomic Validation

NeuroPID was tested with several organisms whose complete proteomes were acquired from UniProt and SwissProt (ranging from 10K for *A. mellifera* to 23K for *C. elegans*). Predictions using just a single model resulted in a large number of weak predictions (ranging from 1.3K to 5.2K candidate NPPs). Including a more stringent probability threshold lead to a drastic reduction in the output, leaving only the top ranked NPP predictions. For example, using a Random Forest model, a probabilistic/'certainty' threshold of 0.99 (i.e. 99% of the ensemble's component trees gave a positive prediction) reduced the unfiltered predictions for *Bombyx mori* from over 4000 to only 16 positive predictions (and 819 negative predictions). Similarly, the application of Gradient Boosting (GBR) for the Monarch butterfly (16183 sequences, 4856 predictions) reduced the number of predictions 10 fold at a 99.5% threshold. Experimental validation will always be the bottleneck, so probabilistic filtration for high-confidence candidates, with a high level of precision, is of obvious use.

Table 4 summarizes the results of NeuroPID predictions in view of annotated NPPs from UniProtKB (partitioned by SW and UniProt). The number of NeuroPID top predictions for most organisms matched the number and identity of curated SW identifications. In all cases, the prediction list matched the list of curated NPPs as compiled in SW (Table 4).

| organism / taxa | # of UniProt (UniRef90) | # of NPP in SW (UniRef90) | # of NPP in UniProt (UniRef90) | Prediction NeuroPID RBF[a] |
|---|---|---|---|---|
| Apis Melliferra | 10394 | 6 | 19 | 7 |
| SP in Apis Mellifera[b] | 2139 | 5 | | 7 |
| Gallus gallus | 20760 | 5 | 5 | 5 |
| SP in Gallus gallus | 701 | 1 | | 1 |
| Bombyx mori | 15250 | 5 | 17 | 9 |
| SP in Bombyx mori | 112 | 5 | 5 | 9 |
| Octopoda | 224 | 4 | 4 | 4 |
| SP in Octopoda | 76 | 3 | | 3 |

Table 4 NeuroPID predictions for individual species. Predicted using SVM with RBF (Radial Basis Function) kernel. SP, Signal peptide predicted according to SignalP 4.0.

### 1.2.6 NeuroPID Discovery protocol

The NPP and NP prediction protocol for NeuroPID is summarized in Figure 9

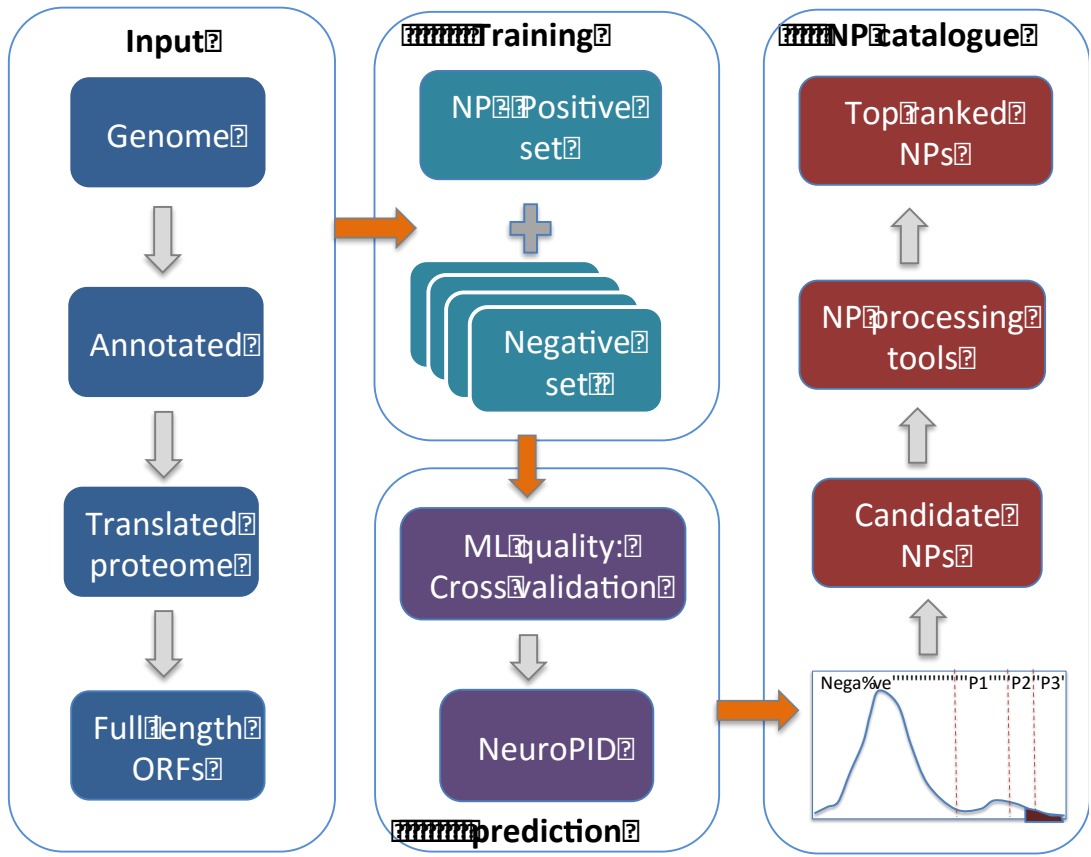

Figure 9 NeuroPID Pipeline. A protocol for NPP prediction using NeuroPID. The protocol is composed of 3 sections: (i) Collecting proteome-scale ORFs; (ii) NeuroPID feature extraction, followed by a trained ML model's prediction (iii) Assessment of the predictions and candidate NPPs.

A refined list of NPPs candidates can be created by prefiltering sequences according to simple criteria, such as the presence of an N'-terminal Signal Peptide, and the absence of transmembrane domains.

NeuroPID was used to identify novel NPPs. In each case, the organism's annotated NPPs were excluded from the model training phase. The NeuroPID results for NPPs from worm, bee, ant and octopus are found in www.protonet.cs.huji.ac.il/neuropid/results/. We analyzed the results for the *Apis mellifera* (Honey bee) and *Thaumeledone gunteri* (Octopoda) proteomes. Table 4 shows the results and the consistancy among the applied methodologies. Consistancy among the decision tree classifiers (RF, GBR) was expecially high. Using all 4 mentioned models, we identified multiple sequences supported by at least 3 of the predictors. (Table 5).

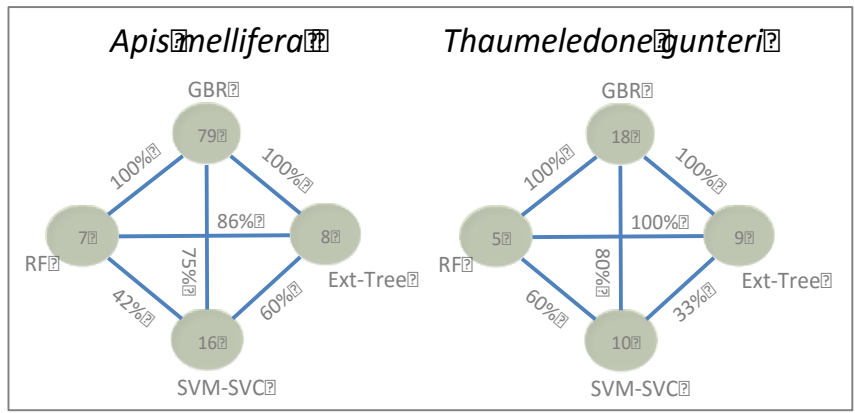

Table 5 NeuroPID predictions for honey bee (Apis) and octopus (Gunteri) proteomes. The number of predictions is shown (nodes) along the % of overlap for the prediction pairs. The % is calculated from the smaller number of predictions in the pair. Random Forest (RF), Extra Trees classifier (Ext-Tree), SVM-SVC (linear kernel) and Gradient boosting trees (GBR).

As an illustration, one of these sequence that was consistently identified by the different ML methods was an uncharacterized sequence from *Apis mellifera (*UniProt: H9K152, 183 amino acids). We propose this sequence to be an overlooked NPP: (i) It is a secreted protein (+SP). (ii) Using MS identification, a 52 amino acid peptide belonging to the full length ORF was identified in the honeybee brain extract (Audsley and Weaver, 2006). (iii) Pfam (Finn *et al.*, 2014) identifies the CRF (Corticotropin-releasing factor) domain, associated with active peptides from frog, fish and mammals. This domain is postulated in stress and anxiety, osmolarity, thermoregulation, growth and metabolism in mammals (Chang and Hsu, 2004).

As part of further ongoing research into Cnidarian "Neuropeptide like peptide-hormones", we extracted a large number of high confidence predictions using the NeuroPID website. We extracted multiple Cnidarian proteomes and predicted ORFs, and ran them through a preliminary pipeline, then NeuroPID. The initial pipeline consisted of using a number of bioinformatics tools to detect "necessary but not sufficient" properties, essential (but not exclusive) to secreted proteins or NPPs: the presence of a signal peptide (predicted using SignalP (Petersen *et al.*, 2011) and Phobius (Käll *et al.*, 2007)) and the lack of a transmembrane or intermembrane region

(detected using Phobius). These secretome candidates were then run though the NeuroPID discovery protocol. The top ranked candidates for each proteome were then manually analyzed and annotated for future experimental validation. The list of high quality predictions is provided in Appendix 4. The manual annotations are unpublished and available upon request.

## 4.3  ProFET Results

ProFET can be used as a baseline for features and model training, or it can be customized with new or specific features. ML models can be trained and saved according to extracted features and annotations. For details about the framework, test-cases and in-depth analyses, we refer to the original article (Ofer and Linial, 2015), included in the appendix.

### 1.2.7  ProFET Benchmarks

Each set was measured using 15 iteration randomized stratified shuffle-split CV. For each iteration, a fraction of the data (18%) is randomly set apart as an evaluation set. The framework automatically selects the best performing classifier. The term "Dummy" applies to a classifier that always picks the majority class. Altogether, we present 17 datasets. For 76% of the datasets accuracy and F1 Scores are above 80%, while for 35% the accuracy is >90% (Figure 10).

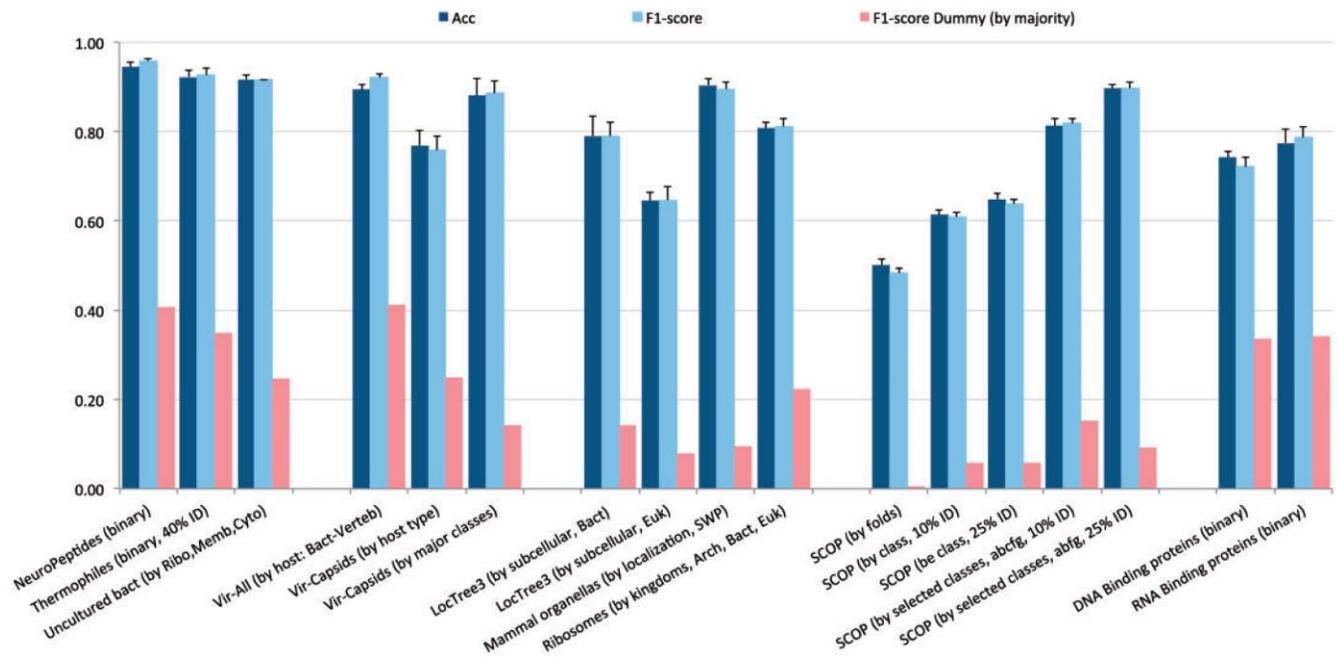

Figure 10 Classification performance by Accuracy and F1 score for 17 datasets using ProFET. Results for Accuracy (Acc, dark blue) and the F1-score (light blue, middle) are shown. Dummy predictor is a default classifier which always predicts the largest class in a dataset (rightmost, pink). Multiple rounds of stratified shuffle-split CV were performed, and the scores were averaged, SD error bars shown.

The classification performance for DNA and RNA binding proteins meets the state of the art results obtained by special purpose predictors (Wang *et al.*, 2010), despite our using a harder (redundancy filtered) version of the data. These specialized predictors for DNA and RNA binding proteins relies on evolutionary information (e.g., PSSMs), requiring computationally intensive alignment of candidate sequences using PSI-BLAST (both for training and inference). On the original, easier dataset, the DNAbinder method reported only 61.42% accuracy and 63.5% using PSSM profiles (Kumar *et al.*, 2007). Our platform reached 72% and 79% accuracyfor DNA and RNA binding proteins, respectively, despite not using evolutionary information. Five of the benchmarks concern structural SCOP datasets, at the class or fold level. We did not attempt to predict secondary structure directly, although this can easily be integrated (e.g. as CTD features), and indeed we do not aim to replace dedicated structural predictors. Naturally, classification success varies by task. For example, the success for the SCOP 'selected class' is very high (0.82–0.9), whereas the performance for the fold classification is much lower (0.62–0.65). In the large SCOP datasets, we added reduced alphabet (Ofer_14) 3-mer frequencies to the features. Note that SCOP 25% and SCOP 10% tasks use the same dataset (SCOPe), differing only by the degree of redundancy removal. We conclude that this excellent performance shows the applicability of the "default" ProFET workflow and features.

### 1.2.8 ProFET Case studies

We selected three datasets for a more detailed analysis. Thermophiles, NPP V2 and Uncultured bacteria. It bears mentioning that the NPP V2 dataset is more challenging than that using in NeuroPID, due to the more challenging negative set consisting of secreted, processed proteins. The uncultured bacteria 'metagenomic' dataset is also challenging given how poorly characterized its progenitors are.

For all three sets, we obtained exceptional classification. Results were derived from K-fold stratified CV (without shuffle-split), no feature selection. Results for the Thermophiles and NPP V2 are shown below in Figure 11. Performance for the uncultured bacteria localization task was also convincing (tested via 12 rounds of stratified shuffle split CV): F1 score was 0.917 (+- 0.01 SD), Accuracy 0.916 (+- 0.01 SD).

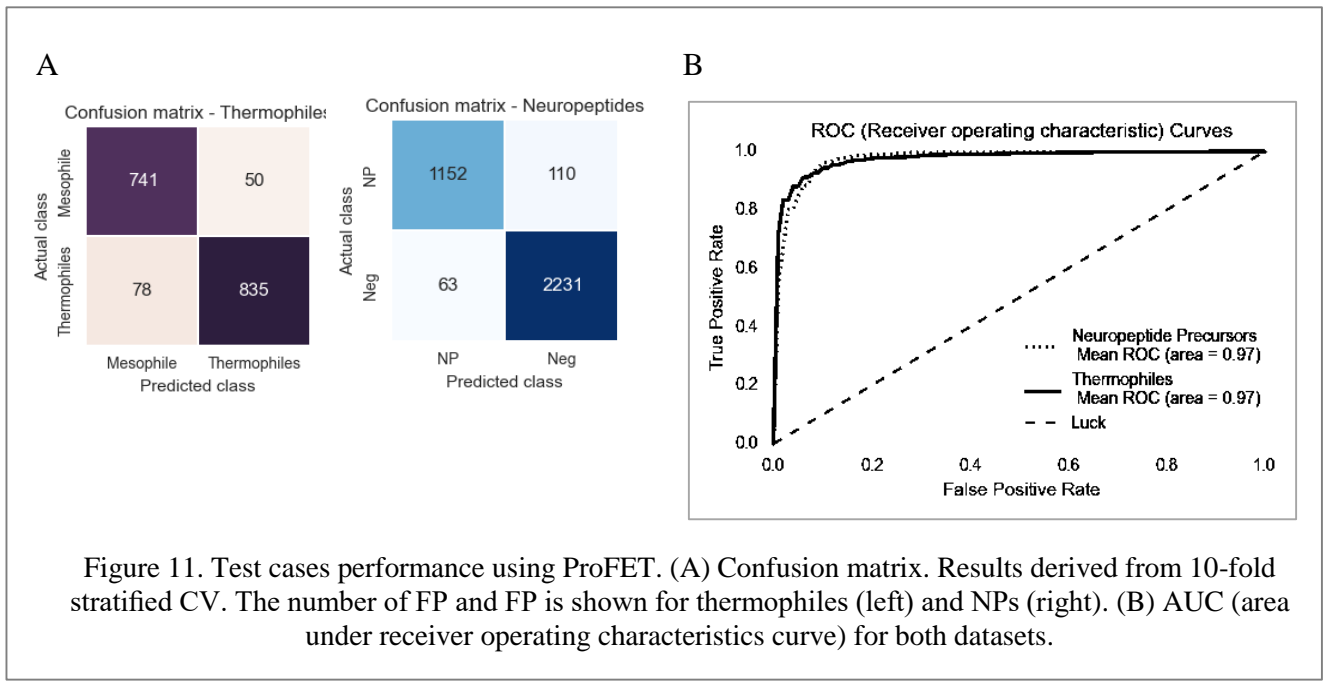

Figure 11. Test cases performance using ProFET. (A) Confusion matrix. Results derived from 10-fold stratified CV. The number of FP and FP is shown for thermophiles (left) and NPs (right). (B) AUC (area under receiver operating characteristics curve) for both datasets.

### 1.2.9 Sequence Alignment Comparison

We compared the most popular sequence alignment method available (PSI-BLAST (Altschul *et al.*, 1997)) to our feature driven approach. The poor performance of sequence alignment methods on structural classes has already been documented (Shah *et al.*, 2008; Ding and Dubchak, 2001), so we chose to compare our novel test cases. This comparison was artificially biased in favor of PSI-BLAST, as we limited its predictions to the training data, rather than the whole protein universe (making the task far easier).

We used PSI-Blast (3 iterations, default parameters) on the Thermophile and NPP V2 sets. The most significant E-value for each sequence was used as an approximate distance matrix. We then trained a K-nearest-neighbors classifier, tested hyperparameters and recorded the best performance. We also used a clustering approach (Spectral clustering, K-means (Yu and Shi, 2003; Moore, 2004)), and compared the clusters to the true labels.

Clustering performance was significant lower than our own (Figure 11). The best results for the Psi-Blast test were obtained from Spectral clustering model. The F1 score was 0.56 and 0.29 for the NPP and Thermophile sets respectively. To make sure the poor performance was not dependent on the choice of ML methodology, we repeated the analysis as a supervised classification task using a K-nearest neighbors classifier (k=1). The data was split 80/20 into evaluation and hold-out sets, and the best hyperparameters were determined by 4 fold CV. For the NPP and Thermophiles sets the accuracy on the 'evaluation set' were 62.8% (+/-0.16 SD) and 48.9% (+/-0.03 SD), respectively. The F1-score was 0.61 and 0.44 respectively. Clearly, pure sequence alignment is inferior to our feature driven approach.

### 1.2.10 ProFET: Feature Selection

Using the RF-RFECV method, we analyzed the top 15 features for the NPP V2 and Thermophile datasets, shown in Figure . The initial set included 771 features.

## Thermophiles informative features:

We note the importance of AA composition, particularly charged and polar amino acid groups. Of notable importance are features involving the organizational entropy of E and Q. The relevance of these AA's organizational entropy in thermophiles had been established (Michelitsch and Weissman, 2000), but I was not aware of this prior to this post-hoc analysis, so this insight was obtained purely from the feature selection. Merely using the AA composition would not have captured many of these features.

The CV performance (F1 score) with just 15 features reached 99.53% of that obtained using all statistically significant features (F1 score=0.906; 453 features).

## NPPs informative features:

As opposed to the features dominating the test case of thermophilic proteins, in the case of the NPPs, a smaller set of features dominates the classification, mainly relating to the normalized frequency of putative dibasic cleavage sites. Further properties of the basic residues Lys (K) and Arg (R) repeat themselves by virtue of entropy, binary autocorrelation (6/15 features) and more. Additional features include protein size (Mw and length) and to a lesser extent some pseudo-structural propensities, such as the B-factor flexibility ('Flex_min'), and secondary structural propensities – reflecting the importance of availability of the putative cleavage sites and atypical composition of the putative peptides. Cysteine (C) importance is expected, given the importance of cysteine cross linkages in short secreted peptides and structurally similar neurotoxins, discussed in ClanTox and our previous work (Tirosh *et al.*, 2013).

The CV performance (F1 score) with just 15 features was 95.85% of that obtained using all statistically significant features (0.945 ($\mp$ 0.01); 544 features).

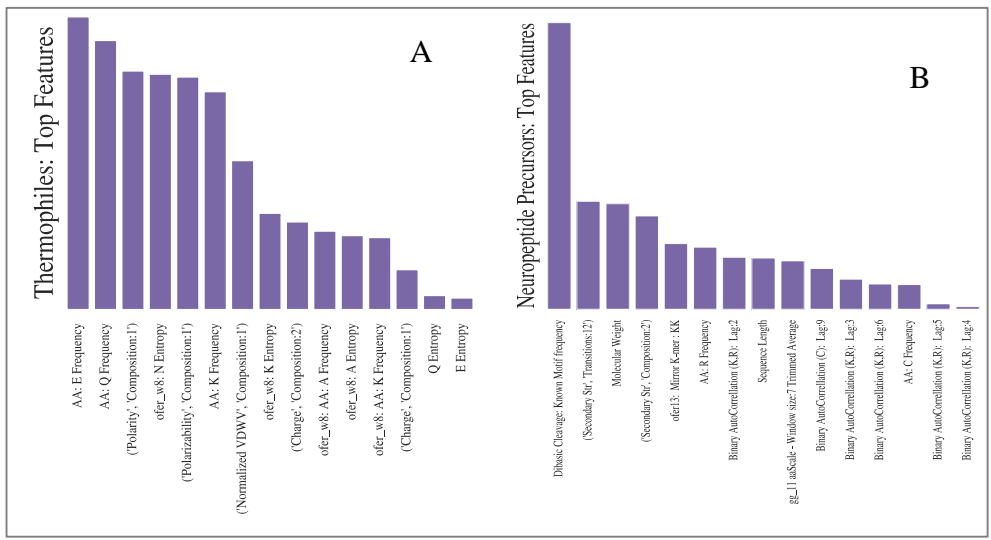

Figure 12 Top 15 features for classification of: (A) Thermophilic vs Mesophilic proteins and (B) NPPs vs non-NPP secreted proteins. Y Axis shows relative feature importance in the model. AA – Amino Acid.

Trimmed average: AA Propensity scale average for a window with the bottom and top 20% extremes removed. 'Scale Name grouping':'Composition 1' – CTD (C) Feature for a given compressed grouping, and one of its 3 subsets. Ofer_w8: 8 letter reduced alphabet grouping.

# 5 DISCUSSION – דיון וסיכום

## 5.1 ProFET Perspectives

Several conclusions can be drawn from our results:

A. Protein centric analysis: The alignment-free feature based methods should be considered a baseline approach for whole proteins, rather than protein domains. Most of our knowledge from 3D structure and evolution relies on the properties of domains within proteins. We propose the feature-centric approach as complementary to the alignment or domain-centric one.

B. "One size fits all": Features in ProFET are highly relevant to a broad range of proteins and classification tasks. This is in contrast to task-specific methods. Therefore, ProFET eliminates duplicated efforts for feature extraction.

C. Flexibility of use: Our presented pipeline accepts a single sequence, combined files, multiple files or a directory. It automatically labels the input into classes (if desired), and transforms the features. Users can set the desired combination of features from ProFET. From the point of view of the user, several consideration were taken:

- Our pipeline handles FASTA files and stores them, annotations, features and predictions.
- We use state of art, free python data science tools (such as Pandas, scikit-learn, biopython)
- Easy to add new features using a standardized format.
- Our framework includes feature details in the pipeline so results are easily interpretable.
- Our code is freely available.

We provide a large collated resource for protein features. Thanks to the modular design of ProFET, adding and tinkering with features is trivial. Users of ProFET can decide to focus, remove or expand any subset of the generated features.

In summary, the approach presented here is suitable and powerful for application towards modern approach for ML especially in the emerging field of Deep Learning, and unsupervised learning of feature representations. These features can be easily experimented with, allowing additional applications of biological insight to the task of feature engineering.

## 5.2 Pitfalls & Tricks

When constructing a dataset for classification, definition of the classes is essential, both from a methodological and practical POV. Biology poses a number of challenges to defining data:

1. Ambiguous definitions.
2. Missing, unlabeled, noisy and missing data.
3. Imbalanced classes.
4. Picking representative sets

These are inherent to virtually any biological problem, and there is no simple solution to any of them. At best, I would note a number of hard-won guidelines and tricks. 1) Care should be taken when using predefined keywords. For example, we found that NP receptors were being included in our initial data, based on the NP keyword. Structured ontologies such as GO (Ashburner *et al.*, 2000) can help, as can gold standard datasets from the literature and the use of manually reviewed data when constructing the initial training set.

2) As in any problem, raw data must be carefully examined then cleaned, using automatic methods. E.g. non-standard amino acids, or non-standard modifications to formats.

3,4) Even the best models will have difficulty when the disparity between groups is extreme. When it's 50,000:1 (1,100 NPPs vs ~56 million), even more so. In our initial work we addressed this using lengthwise stratified sampling generate a balanced negative set. We noted that this gave better generalization performance than supermajority sets, even when repeated multiple times. A second trick is stringent redundancy removal on the majority class – this maximizes diversity, albeit at the cost of making the task more difficult (an acceptable compromise). A further improvement utilized in the NPP V2 dataset was to "prefilter" our negative set using biological knowledge: Rather than sampling from the (100%) whole protein universe, we sampled from among the proteins most like our positive class (+SP,-TMD and same range of lengths). Given there are only 3.4 million such proteins, we sample from just ~8% of the proteome. This gives a set which is far more similar to our "positives", but one more representative of the input we expect from real world users. We did not even need to sample randomly from the negative set to get a "balanced" ratio of samples. Making the negative set "harder" is a common theme in our work, with the goal of ensuring the usefulness of our models. I note that care must be taken when looking at an unknown background for "negatives" (Charles Elkan), as many unlabeled sequences could be positives. Even more so, non-representative selections are hazardous: In NeuroPID we used proteins with SP and TMD: Proteins with SP meant we were looking at just ~10% of the proteome, and proteins with a TMD represented a large and varied set, that could not be NPs. CV performance was very high, but later examination on hold-out proteins revealed that the classifier had not learned to identify between Neuropeptides and non-NPs, but rather to identify proteins with a TMD! (i.e the TMD was too easy and distinct a feature).

## 5.3 ML and Predictive Proteomics

As noted, the main drawbacks in existing sequence-based methods are (i) Some functions cannot be detected by sequence-based methods; (ii) Current statistical models mostly capture local patterns rather than high-level function and (iii) rare sequences or properties with few homologs cannot be successfully used for inference or construction of good statistical model via MSA. Furthermore, inference on new sequences is growing ever more computationally costly, since sequence alignment methods rely on all vs all searching. Even the best clustering methods can only do so much to reduce the search space from tens of millions of sequences, to a few million (Rappoport and Linial, 2013; Hauser *et al.*, 2013), and this problem is only growing. Alignment free features allow rapid inference for new samples, without needing to search anew through the existing data. This is in addition to their greater flexibility, ability to learn high level features and correlations, and the ability to integrate multiple levels of features and information.
However, such ML approaches are still best suited to certain use cases, such as those raised here, e.g. high level function, cases lacking structural or sequence similarity, or searching proteins for a specific novel function. However, in the "naïve" use case of annotating highly similar sequences, "guilt by association" and alignment is likely to remain the method of choice for the foreseeable future, due to its precision in the cases where it works best. Other use cases are a different matter however, and an ever increasing number of computational methods integrate ML in their approaches (Radivojac *et al.*, 2013; Kryshtafovych *et al.*, 2014), albeit in a huge variety of ways, and variety of feature representations (e.g. most integrate PSSM profiles, or existing ontological annotations rather than sequence-derived alignment free features).

## Deep learning

So called "Deep learning" ML methods are variants of artificial neural networks (NN) with multiple hidden layers and domain-specific architectures (such as convolutional NN or recurrent NNs (RNN)). These have attained state of the art performance in a wide range of ML domains in recent years, notably in computer vision and natural language processing (Hochreiter and Schmidhuber, 1997; Xu *et al.*, 2015; Kim, 2014; Krizhevsky *et al.*, 2012; Simonyan and Zisserman, 2014; LeCun *et al.*, 2015). It is likely that the next step in improve computational protein function and family prediction will use these powerful models, although so far most have been used only for residue level prediction (Lyons *et al.*, 2014; Zhou and Troyanskaya, 2014; Qi *et al.*, 2012), rather than whole sequences (Hochreiter *et al.*, 2007). This stems mainly from two problems: 1) the difficulty of representing whole sequences in a fixed length form (Nanni *et al.*, 2014). This could be overcome using RNNs, distributed representations (in the manner of Word2Vec) (Le and Mikolov, 2014; Goldberg and Levy, 2014; Asgari and Mofrad, 2015) or transformation of aa propensity scale representations into a fixed length time series. Little of this has been done, although I am currently pursuing this as part of future research.

2) The ability of NNs to universally approximate any function is also their Achilles heel: being very high variance methods, they require a large amount of training samples and are vulnerable to overfitting (Geman *et al.*, 1992). Other lower variance ML methods may be less flexible, but are also less prone to overfitting, while statistical methods such as HMMs or Conditional Random fields can learn more efficiently than "vanilla" NNs. In the domain of computer vision, a successful solution is transfer learning (Yosinski *et al.*, 2014; Sharif *et al.*, 2014): Given a specific domain, train a large model and learned feature representation (e.g. via convolutions or embeddings (Mikolov *et al.*, 2013)) on a challenging task with universal commonalities and a large diverse dataset (e.g. Imagenet). The feature representations may then be used to extract features in other tasks. This does not currently exist in Bioinformatics, where both data and class definitions are hard to define and gather. I do propose that such a transfer learning model for some tasks reliant on structural homology could be created using the large protein family datasets gather in SCOP, CATH or PFAM, combined with our own set of alignment free features, autoencoders (Vincent *et al.*, 2008; Xu *et al.*, 2011), distributed representations of proteins as sentences (Le and Mikolov, 2014), and convolutional RNNs (Sønderby *et al.*, 2015), but this is beyond the scope of this work.

That said, we experimented with deep learning (multi-layer perceptrons) during this work and did not obtain significantly better results (unpublished). An oft ignored aspect is one of cost-benefit: methods such as RF are very fast, handle class imbalance gracefully, are robust to noisy features and high features dimensionality, do not need special input scaling, allow feature selection, and are very easy to use with few hyperparameters. This made them a far more practical choice, even more so given the varied, imbalanced datasets we faced.

## 5.4 Overall Conclusions

A curated set of best practices is essential to preventing scientists from continually "reinventing the wheel" (Ofer and Linial, 2015; Li *et al.*, 2006). This would lower the barrier of entry to new researchers, allowing an influx of novel, interdisciplinary ideas from domain experts (e.g. experts on molecular toxicology). A predefined set of good, "generic" features is beneficial to getting a "good enough" baseline for different tasks, as shown in ProFET.

ProFET's performance was assessed rigorously and reached state of the art results in the vast majority of tasks. Our "default" set of features performed well in all the tasks, even when compared to existing, specialized predictors. I introduced many novel features. For example, features based on reduced alphabets (e.g. K-mers, CTD), entropy, high performance AA scales and their usage, binary autocorrelation, sequence segmentation, mirror k-mers and more. Many of these features not only improved performance while allowing a compact representation, and exposing important biologically interpretable properties (Figure ).

Profet is flexible, universally applicable to any protein sequence, was rigorously tested, is easy to use, and has state of the art "out of the box" performance.

Our Neuropeptide identification platform is the first of its kind, and offers excellent performance and scalability. Its availability as a webtool makes it a convenient tool for research into new neuropeptides. We discovered many putative NPs, and hope to enable future discoveries into peptide regulators of behavior and metabolism based on the platform and discovered NPs.

# 7 APPENDICES

# APPENDIX 1 NEUROPID: A PREDICTOR FOR IDENTIFYING NEUROPEPTIDE PRECURSORS FROM METAZOAN PROTEOMES

*Sequence Analysis*

## NeuroPID: A Predictor for Identifying Neuropeptide Precursors from Metazoan Proteomes

Dan Ofer[1] and Michal Linial[1,*]

[1] Department of Biological Chemistry, Institute of Life Sciences, The Edmond J. Safra Campus, The Hebrew University of Jerusalem, Givat Ram, 91904, Israel

Associate Editor: Dr. John Hancock

**ABSTRACT**

**Motivation:** The evolution of multicellular organisms is associated with increasing variability of molecules governing behavioral and physiological states. This is often achieved by Neuropeptides (NPs) that are produced in neurons from a longer protein, named neuropeptide precursor (NPP). The maturation of NPs occurs through a sequence of proteolytic cleavages. The difficulty in identifying NPPs is a consequence of their diversity and the lack of applicable sequence similarity among the short, functionally related NPs.

**Results:** Herein, we describe NeuroPID, a machine-learning scheme that predicts Metazoan NPPs. NeuroPID was trained on hundreds of identified NPPs from the UniProtKB database. Some 600 features were extracted from the primary sequences and processed using support vector machines (SVM) and ensemble decision tree classifiers. These features combined biophysical, chemical and informational-statistical properties of NPs and NPPs. Other features were guided by the defining characteristics of the dibasic cleavage sites motif. NeuroPID reached 89-94% accuracy and 90-93% precision in cross-validation blind tests against known NPPs (with an emphasis on Chordata and Arthropoda). NeuroPID also identified NPP-like proteins from extensively studied model organisms as well as from poorly annotated proteomes. We then focused on the most significant sets of features that contribute to the success of the classifiers. We propose that NPPs are attractive targets for investigating and modulating behavior, metabolism and homeostasis, and that a rich repertoire of NPs remains to be identified.

**Availability:** NeuroPID source code is freely available at http://www.protonet.cs.huji.ac.il/neuropid

**Contact:** michall@cc.huji.ac.il

## 1 INTRODUCTION

Peptides are known to be key modulators in behavior, sensation and homeostasis (Brain and Cox, 2006). Biologically active peptides that are produced and secreted from neurons and act to modulate their function are collectively called peptide modulators or Neuropeptides (NPs). NPs represent a widespread mode of communication that is found from Cnidarians to Bilaterians, including mammals. NP precursors (NPPs) are subjected to regulated cleavages that result in the production of functionally active NPs. The processing of NPs occurs along the secretory pathway (Gelman and Fricker, 2010). In most instances, NPs locally modulate pre-synaptic or postsynaptic cell activity (Southey, *et al.*, 2008). From a functional perspective, known effects of NPs include stress control, pain perception, social behaviors and sleep-wake cycle (Nassel, 2002). NPs also regulate food uptake, maintaining appetite and body weight. The peripheral NPs that act outside the central nervous system (CNS) regulate gastrointestinal and immunological functions (Insel and Young, 2000). For example, some NPs induce chemotactic response in attracting immature dendritic cells to the site of inflammation. In this context, substances P and K, two extensively studied NPs, induce the production and release of inflammatory cytokines from blood monocytes (Gonzalez-Rey, *et al.*, 2007). "Social" NPs such as oxytocin (OXT) and arginine vasopressin (AVP) regulate complex social cognition and behavior (including pair-bonding, social recognition and maternal behavior) (Insel and Young, 2000). The immense diversity among NPs contributes to the wide range of behavioral tasks that are carried out.

A common feature for the majority of NPs is their production from a larger precursor (NPP). The production of short bioactive peptides is a result of a series of cleavages and maturation events (Mirabeau, *et al.*, 2007). NPP can produce multiple copies of different NPs (Mentlein and Dahms, 1994). Notably, a cluster of basic residues specifies these cleavage sites. The occurrence of dibasic residues specifies the canonical sites for intracellular endopeptidases such as Furin (Veenstra, 2000). However, some NPs that act in cell-cell communication (Funkelstein, *et al.*, 2010) (*e.g.*, cathepsin L) do not obey the dibasic residues specificity rule. The identity and regulation of the key peptidases that are responsible for the processing of the NPP remains an active research field.

NPs typically bind cell surface G-proteins coupled receptors (GPCRs) that initiate a signaling cascade. An evolutionary analysis of NPPs and their cognate GPCRs suggested a diversification that occurred in certain taxonomical branches (Jekely, 2013). The NPs and their receptors are attractive targets for drug development and translational medicine based on their role in feeding, sexual behavior and cellular homeostasis (Brain and Cox, 2006).

The goal of this research is to enable systematic identification of NPPs (and NPs) at a genome-wide scale. The difficulty in identifying NPs and classifying genes as potential NPs stems from the following: (i) Current gene annotation tools mostly rely on se-

*To whom correspondence should be addressed.

# APPENDIX 2 NEUROPID: A CLASSIFIER OF NEUROPEPTIDE PRECURSORS

# NeuroPID: a classifier of neuropeptide precursors

Solange Karsenty[1,2], Nadav Rappoport[1], Dan Ofer[3], Adva Zair[1] and Michal Linial[3,*]

[1]School of Computer Science and Engineering, The Hebrew University of Jerusalem, Jerusalem, Israel, [2]School of Computer Science, Hadassah Academic College, Jerusalem, Israel and [3]Department of Biological Chemistry, The Alexander Silberman Institute of Life Sciences, The Sudarsky Center for Computational Biology, The Hebrew University of Jerusalem, Jerusalem, Israel

## ABSTRACT

Neuropeptides (NPs) are short secreted peptides produced in neurons. NPs act by activating signaling cascades governing broad functions such as metabolism, sensation and behavior throughout the animal kingdom. NPs are the products of multistep processing of longer proteins, the NP precursors (NPPs). We present NeuroPID (Neuropeptide Precursor Identifier), an online machine-learning tool that identifies metazoan NPPs. NeuroPID was trained on 1418 NPPs annotated as such by UniProtKB. A large number of sequence-based features were extracted for each sequence with the goal of capturing the biophysical and informational-statistical properties that distinguish NPPs from other proteins. Training several machine-learning models, including support vector machines and ensemble decision trees, led to high accuracy (89–94%) and precision (90–93%) in cross-validation tests. For inputs of thousands of unseen sequences, the tool provides a ranked list of high quality predictions based on the results of four machine-learning classifiers. The output reveals many uncharacterized NPPs and secreted cell modulators that are rich in potential cleavage sites. NeuroPID is a discovery and a prediction tool that can be used to identify NPPs from unannotated transcriptomes and mass spectrometry experiments. NeuroPID predicted sequences are attractive targets for investigating behavior, physiology and cell modulation. The NeuroPID web tool is available at http://neuropid.cs.huji.ac.il.

## INTRODUCTION

The functions elicited by neuropeptides (NPs) cover most aspects of metazoan life including metabolism, growth and social behavior (1). For example, in mollusks and insects, mating behavior and reproduction are regulated by NPs. More generally, NPs function in energy consumption, food uptake, pain and sensation, temperature control, appetite, mating and social behavior. NPs act through binding their cognate receptor and activating a signaling cascade (2).

Most mature NPs are short peptides (5–30 aa) that are produced from a longer polypeptide, referred to as a NP precursor (NPP). Active NPs are the products of multiple cleavages of the precursor. Such cleavages mostly occur at basic residues (Arg and Lys) motifs that flank the NPs. While dibasic residues are the hallmark of the cleavage sites for endopeptidases (3), in some NPs such canonical cleavage sites are not detected.

Several bioinformatics approaches were developed to enhance the routine BLAST searching protocol by incorporating distinctive properties of NPs (e.g. (4)). However, at present, there is no systematic approach to identify NPPs in poorly annotated metazoan genomes. The organization of the NPP as a source of multiple NPs is conserved despite a weak similarity between NPPs along the evolutionary tree (5). Currently, NPPs are sporadically identified from sequencing projects of organisms and from the collections of large-scale proteomics and transcriptomics experiments. We developed a systematic approach for identifying candidate NPPs (6). Large-scale mass spectrometry (MS) proteomics identified thousands of spectra that derived from NPP genes. Additionally, a large number of bioactive modulatory peptides were identified (e.g. (7)). For peptides collected from human samples, only about 20% of the associated genes are currently annotated as NPPs. The expansion in MS proteomics and the depth of the transcriptome coverage suggest that NeuroPID is valuable as a discovery platform for bioactive modulators. NPPs and their associated NPs are attractive targets for drug development and pharmacological manipulations.

## OUTLINE—FEATURES TO PREDICTIONS

Figure 1 shows a prototype of an NPP that belongs to the Allatostatin family as an illustration for the challenges in identifying such genes at a genomic scale. The Allatostatins are produced in many insects to control food intake (8). The sequence of the Pacific beetle cockroach (Figure 1) contains 13 identified NPs. While each peptide has a unique se-

*To whom correspondence should be addressed. Tel: +972 2 6585425; Fax: +972 2 6586448; Email: michall@cc.huji.ac.il

# APPENDIX 3 PROFET: FEATURE ENGINEERING CAPTURES HIGH-LEVEL PROTEIN FUNCTIONS

Sequence analysis

## ProFET: Feature engineering captures high-level protein functions

Dan Ofer and Michal Linial*

Department of Biological Chemistry, Institute of Life Sciences, The Edmond J. Safra Campus, The Hebrew University of Jerusalem, Givat Ram, 91904, Israel

*To whom correspondence should be addressed.
Associate Editor: John Hancock

**Abstract**

**Motivation:** The amount of sequenced genomes and proteins is growing at an unprecedented pace. Unfortunately, manual curation and functional knowledge lag behind. Homologous inference often fails at labeling proteins with diverse functions and broad classes. Thus, identifying high-level protein functionality remains challenging. We hypothesize that a universal feature engineering approach can yield classification of high-level functions and unified properties when combined with machine learning approaches, without requiring external databases or alignment.
**Results:** In this study, we present a novel bioinformatics toolkit called ProFET (Protein Feature Engineering Toolkit). ProFET extracts hundreds of features covering the elementary biophysical and sequence derived attributes. Most features capture statistically informative patterns. In addition, different representations of sequences and the amino acids alphabet provide a compact, compressed set of features. The results from ProFET were incorporated in data analysis pipelines, implemented in python and adapted for multi-genome scale analysis. ProFET was applied on 17 established and novel protein benchmark datasets involving classification for a variety of binary and multi-class tasks. The results show state of the art performance. The extracted features' show excellent biological interpretability. The success of ProFET applies to a wide range of high-level functions such as subcellular localization, structural classes and proteins with unique functional properties (e.g. neuropeptide precursors, thermophilic and nucleic acid binding). ProFET allows easy, universal discovery of new target proteins, as well as understanding the features underlying different high-level protein functions.
**Availability and implementation:** ProFET source code and the datasets used are freely available at https://github.com/ddofer/ProFET.
**Contact:** michall@cc.huji.ac.il
**Supplementary information:** Supplementary data are available at *Bioinformatics* online.

## 1 Introduction

The most used approaches in protein classification rely on distance measures between sequences based on various alignment methods (e.g. Smith-Waterman, BLAST). With the growth in the amounts and diversity of protein sequences, more sophisticated methods have been introduced (e.g. PSSM, Profile-Profile, HMM-HMM) (Jaakkola *et al.*, 2000; Soding, 2005). These methods are based on multiple sequence alignments for improving remote homologs detection (Edgar and Sjolander, 2004; Karplus *et al.*, 1998). Incorporating 3D-structure as a seed for the statistical models further improved the quality of protein domains and families (e.g. Pfam) (Finn *et al.*, 2014; Sonnhammer *et al.*, 1997). Currently, there are ∼27 000 such models (InterPro, Mulder and Apweiler, 2007) that cover 83% of all sequences in UniProtKB (2014_10). Function assignment is gained from mapping InterPro models to Gene Ontologies (i.e. InterPro2GO). An alternative model-free approach

# APPENDIX 4: NEUROPRED SUPPLEMENTARY: STATISTICAL SIGNIFICANCE ANALYSIS OF FEATURES: NEUROPEPTIDES VS BACKGROUND

The following graphs are from the supplementary material of the original NeuroPred article. They reflect differences between neuropeptides (NP and NPPs) and the complete "background" distribution of proteins drawn from Uniprot/Uniref (this set includes the NPs within itself).

## Statistical Significance of Features Analysis: Neuropeptides vs Background:

Elaborations of various features: http://web.expasy.org/protparam/protparam-doc.html#ref7 .

| Feature | K-S Test P Value | T Test P Value | Graph/Plots |
|---|---|---|---|
| Molecular Weight | P = 1.7872e-76 | P = 4.3665e-09 | 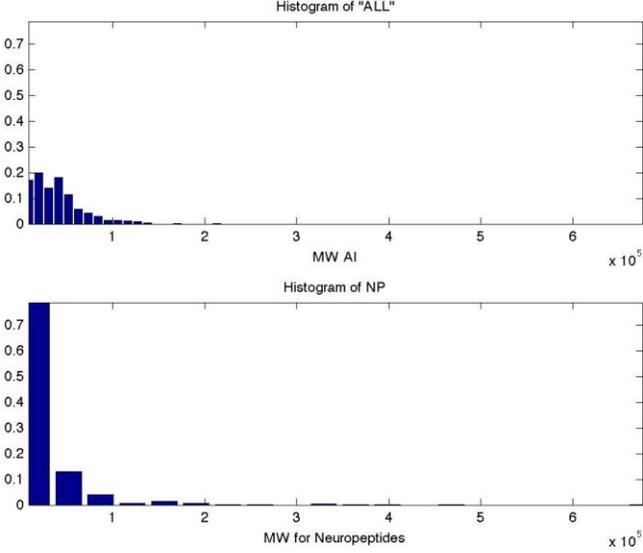 |
| PI | 8.0239e-21 | 1.0797e-16 | 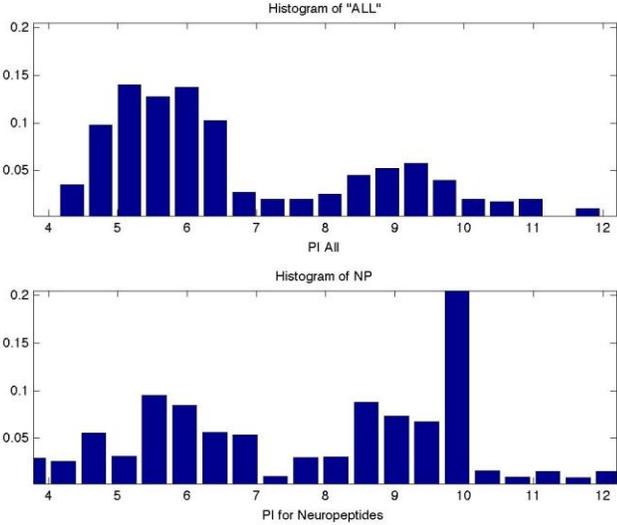 |
| Instability. (A value above 40 means the protein is unstable =short half-life) | 8.6665e-11 | 0.3521 | 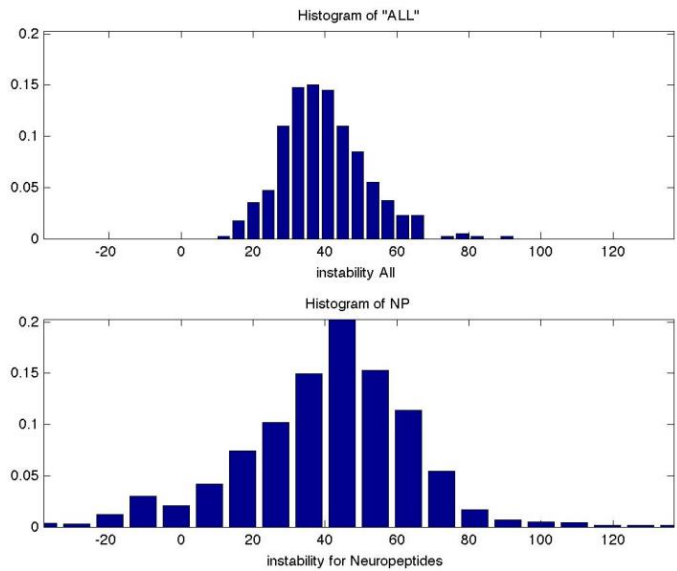 |
| GRAVY | 1.4997e-16 | P = 0.4130 | 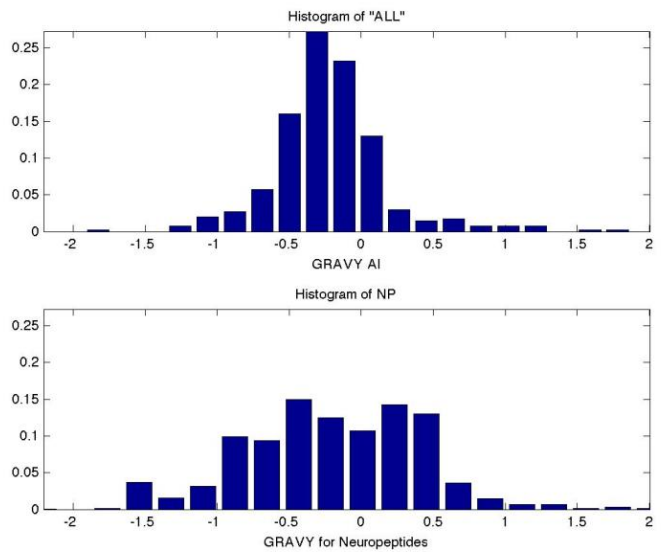 |

| Aromaticity= frequency of Phe+Trp+Tyr | 2.2365e-40 | P = 2.8565e-28 | 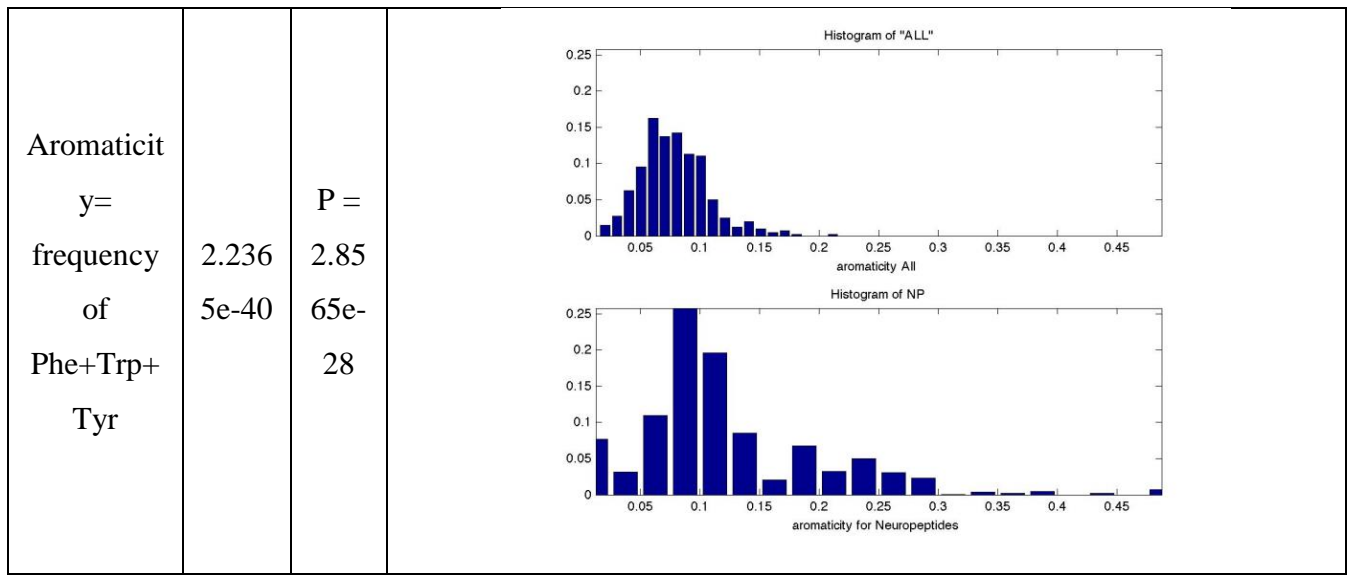 |
|---|---|---|---|

**Amino Acid Frequencies** (NP vs background proteome).

| AA | A | C | E | D | G | F | I | H | K |
|---|---|---|---|---|---|---|---|---|---|
| 2TTest P | 2.8364 e-24 | 0.7209 | 1.4261e-45 | 1.5434e-07 | 5.1228e-22 | 6.6598e-41 | 1.7323e-45 | 0.0431 | 3.9703e-81 |

| AA | M | L | N | Q | P | S | R | T | W | V | Y |
|---|---|---|---|---|---|---|---|---|---|---|---|
| 2TTest P | 1.1661e-15 | 0.0103 | 0.2491 | 0.8762 | 4.8877e-34 | 7.8457e-22 | 3.7566e-08 | 4.5153e-16 | 8.6019e-10 | 3.7784e-26 | 0.7476 |

# APPENDIX 5 IN-SILICO CNIDARIA PREDICTIONS

The full list of predicted, high confidence Neuropeptide-like candidates in a variety of Cnidarians, including Nematostella Vectensis, Hydra Vulgaris, Ctenophora-Sea Gooseberry, and Mnemiopsis, is available on request.

68 of the highest confidence predictions for Nematostella vectensis were manually annotated, and their headers are provided here. The manual annotations themselves are available upon request, as they are part of an ongoing, unpublished research-collaboration. These are Nematostella Vectensis proteins with no known domains in PFAM, a predicted Signal peptide (according to SignalP and Phobius) and no predicted Transmembrane domain, and that are predicted to be Neuropeptides with a high confidence (majority vote by the 4 classifiers) by the online NeuroPID platform.